\documentclass[journal,onecolumn,10pt]{IEEEtran}
\IEEEoverridecommandlockouts

\usepackage{cite}
\usepackage{amsmath,amssymb,amsfonts}
\usepackage[noend]{algpseudocode}
\usepackage{amsthm}

\usepackage{graphicx}
\usepackage{float}
\usepackage{threeparttable}
\usepackage{subfigure}
\usepackage[ruled,linesnumbered]{algorithm2e}
\usepackage{algpseudocode}
\usepackage{textcomp}
\usepackage{xcolor}
 \usepackage{makecell}
\usepackage{pdfpages}
\usepackage{footnote}
\usepackage{diagbox}
\usepackage{bbm}
\usepackage{dsfont}

\newtheorem{theorem}{Theorem}
\newtheorem{Remark}{Remark}

\newtheorem{Definition}{Definition}

\newtheorem{lemma}{Lemma}
\newtheorem{Proposition}{Proposition}
\usepackage{comment}
\usepackage{booktabs,color,amssymb,amsmath}

\def\BibTeX{{\rm B\kern-.05em{\sc i\kern-.025em b}\kern-.08em
    T\kern-.1667em\lower.7ex\hbox{E}\kern-.125emX}}
    
\begin{document}

\title{ {Differentially Private Secure Multiplication with Erasures and Adversaries}

\author{Haoyang Hu,
Viveck R. Cadambe}

\thanks{
Haoyang Hu and Viveck R. Cadambe are with the School of Electrical and Computer Engineering, Georgia Institute of Technology, Atlanta, GA, 30332 USA. (E-mail: \{haoyang.hu, viveck\}@gatech.edu)
This work is partially funded by the National Science Foundation under Grant CIF 2506573.
This manuscript is the full version of our ISIT 2025 paper \cite{hu2025}.
}

}

\maketitle

\begin{abstract}
We consider a private distributed multiplication problem involving $\mathsf{N}$ computation nodes and $\mathsf{T}$ colluding nodes. Shamir's secret sharing algorithm provides perfect information-theoretic privacy, while requiring an honest majority, i.e., $\mathsf{N} \ge 2\mathsf{T}+1$.
Recent work has investigated approximate computation and characterized the tight privacy-accuracy trade-off for the honest minority setting $(N \leq 2\mathsf{T})$ for real-valued data, quantifying privacy leakage via the differential privacy (DP) framework and accuracy via the mean squared error. However, it does not incorporate the error correction capabilities of Shamir's secret-sharing algorithm. 
 This paper develops a new polynomial-based coding scheme for secure multiplication with an honest minority, and characterizes its achievable privacy-utility tradeoff, showing that the tradeoff can approach the converse bound as closely as desired. Unlike the previous scheme, the proposed scheme inherits the capability of the Reed-Solomon (RS) code to tolerate erasures and adversaries.
We utilize a modified Berlekamp–Welch algorithm over the real number field to detect adversarial nodes. 
\end{abstract}

\section{Introduction}
Secure multi-party computation allows multiple parties to collaboratively perform a computation over their private inputs while preserving the confidentiality of those inputs\cite{goldreich1998secure}. 
A key coding theoretic technique in this area is Shamir's secret sharing \cite{shamir1979share}, a method rooted in Reed-Solomon (RS) codes \cite{reed1960polynomial} that provides a framework for ensuring information-theoretic privacy. 
Consider an $\mathsf{N}$-node secure computation system designed to compute the product of two random variables $A, B \in \mathbb{F}$, where $\mathbb{F}$ is a finite field. 
Let $\{R_t, S_t\}_{t=1}^{\mathsf{T}}$ be independent random noise variables uniformly distributed over the field. Shamir's secret sharing algorithm encodes $A, B$ by constructing polynomials:
\begin{align}
    p_1(x)=A+\sum_{t=1}^\mathsf{T} R_t x^t, \quad p_2(x)=B+\sum_{t=1}^\mathsf{T} S_t x^t.    
\end{align}
Node $i$ receives $p_1(x_i)$ and $p_2(x_i)$, where $\{x_1,\cdots,x_\mathsf{N}\}$ denote non-zero distinct elements over the field $\mathbb{F}$.
Any set of $\mathsf{T}$ colluding nodes fail to recover $A$ and $B$, and at least $2\mathsf{T}+1$ nodes are required to obtain the product $AB$ by interpolating the $2\mathsf{T}$-degree polynomial $p_1(x)p_2(x)$.
In other words, perfect information-theoretic privacy requires an honest majority to ensure security and correctness. 
Secure computation techniques (such as the celebrated BGW algorithm \cite{wigderson1988completeness}) use Shamir's secret sharing to enable computation of general polynomial functions (extending beyond basic product computations) over a system of $\mathsf{N}$ nodes such that the information obtained at any $\mathsf{T}$ nodes is statistically independent of the input data, and strictly require $\mathsf{T} < \mathsf{N}/2.$  
Furthermore, because of the error-correction properties of RS codes, secure computation schemes based on Shamir's secret sharing typically tolerate node erasures and adversarial nodes that send erroneous values, as long as the number of erasures and adversarial nodes are bounded by certain thresholds.

In several practical applications, especially in machine learning, a controlled amount of privacy leakage is acceptable, rather than enforcing \emph{perfect} information-theoretic privacy\cite{mcmahan2018general}. Differential Privacy (DP) offers a widely used framework for quantifying and managing this leakage \cite{dwork2006differential}. 
Recent work \cite{cadambe2023differentially} has explored secure multiplication over the real field and within the honest minority setting, i.e., $\mathsf{T}<\mathsf{N}<2\mathsf{T}+1$, and established a tight privacy-accuracy tradeoff through a DP perspective.
Technically, \cite{cadambe2023differentially} introduces multiple layers of noise to the inputs, aiming to achieve DP under $\mathsf{T}$-node collusion while improving the accuracy of estimates. However, the design in \cite{cadambe2023differentially} is divorced from the polynomial structure of RS codes and Shamir secret sharing. Notably, the scheme of \cite{cadambe2023differentially} is asymmetric, requiring one node to implement a unique noise structure distinct from the others. This dependency creates a critical vulnerability, as, unlike RS codes, successful decoding cannot be assured if the designated node either fails (acting as an erasure) or if a subset of nodes behaves maliciously (acting as adversaries).
This prompts the exploration of whether a polynomial-based coding scheme can be developed for the honest minority setting. Such a scheme would aim to provide input data privacy while preserving the inherent robustness of RS codes in tolerating both erasures and adversarial actions. This is the main contribution of this paper.

We develop a novel polynomial-based coding scheme over the real field for differentially private secure multiplication in the presence of erasures and adversaries. 
We characterize the achievable privacy-accuracy tradeoff of our scheme, with privacy quantified by the DP parameter and accuracy quantified by the mean square error. 
Our scheme achieves performance comparable to that of \cite{cadambe2023differentially}, while offering resilience against erasers and adversaries. Specifically, our analysis shows that the privacy-accuracy trade-off achieved by our scheme asymptotically approaches the converse bound established in \cite{cadambe2023differentially}, which is derived under the setting without erasures or adversaries.
We operate within the real domain and use a modified version of the Berlekamp–Welch algorithm \cite{welch1986error} to detect and correct adversaries. 
We verify the correctness of the algorithm via both theoretical analysis and numerical simulations.

\emph{Related works:}
Coding techniques are widely employed in distributed systems not only to improve resilience against stragglers by introducing structured redundancy \cite{lee2017speeding,yu2017polynomial,tandon2017gradient,wan2021distributed,khalesi2022multi} but also to ensure the privacy of sensitive inputs \cite{yu2019lagrange, d2020gasp, jahani2023swiftagg+, akbari2021secure, chang2018capacity, jia2021capacity, liang2024privacy, soleymani2021coded}.
These works extend the standard Shamir's secret-sharing algorithm by introducing additional constraints on the distributed systems, and develop novel coding schemes that ensure both exact computation and perfect information-theoretic privacy.
{
Prior research has primarily focused on finite fields. In contrast, working over the real domain reduces redundancy by allowing for approximation under natural and tractable metrics for both accuracy and privacy. Moreover, in practical applications, quantization from the real domain to a finite field can incur accuracy loss if the conversion is not done carefully. To mitigate these limitations, recent works \cite{soleymani2021analog, soleymani2022analog, liu2023analog, liu2023differentially} have explored distributed computing approaches directly in the real domain, where strict information-theoretic privacy guarantees are generally unattainable.}
{
Specifically, \cite{liu2023analog, liu2023differentially} explore the application of DP to analyze the privacy-utility tradeoff in multiparty computation over real fields, similar to this work. Their schemes are based on complex-valued Shamir's secret sharing, while \cite{cadambe2023differentially} demonstrates that this approach fails to achieve the optimal privacy-accuracy tradeoff characterized in \cite{cadambe2023differentially}.
}
{In addition, error correction codes over the real field are explored in \cite{roth2020analog, jiang2024analog}, where errors exceeding a certain threshold are detected and noise with small amplitude can be tolerated. 
This work also shows that the proposed error decoding algorithm is capable of detecting errors exceeding a certain threshold with probability
approaching 1, while errors below this threshold do not compromise the accurate recovery of the multiplication results.
}
\cite{soleymani2022approxifer} employs the Berlekamp–Welch algorithm \cite{welch1986error} to address the adversarial challenge associated with polynomial-based codes in the real domain. 
In our work, we adopt a modified Berlekamp–Welch algorithm that differs from that in \cite{soleymani2022approxifer} to mitigate the impact of adversaries (detailed in Section \ref{sec::error_correction}).

\emph{Notations}: 
Calligraphic symbols denote sets, bold symbols denote matrices and vectors, and sans-serif symbols denote system parameters.
 $\mathbf{1}$ denotes an all-ones column vector.
For a positive integer $a$, we let $[a]\triangleq \{1,\ldots,a\}$.
For a matrix $\mathbf{A}$, let $\mathbf{A}^T$ denote its transpose. 
For functions $f$ and $g$ and for all large enough values of $x$, we write $f(x)=O(g(x))$ if there exists a positive real number $M$ and a real number $a_0 \in \mathbb{R}$ such that
$|f(x)|\le M |g(x)|$ for all $x\ge a_0$.

\section{System Model and Main Results}
In this section, we introduce the system model and present the main results of this paper. 

\subsection{System Model}
We consider a distributed computing system in which $\mathsf{N}$ nodes collaboratively perform the multiplication of two input random variables $A, B \in \mathbb{R}$.
These random variables are assumed to be statistically independent and satisfy the constraints that $\mathbb{E}[A^2], \mathbb{E}[B^2] \le \eta$ with $\eta \ge 0$.

Let each node $i\in[\mathsf{N}]$ store a noisy version of inputs $A$ and $B$ as follows: 
\begin{subequations}
    \begin{align}
        &\tilde{A}_i=A+\tilde{R}_i, \\
        &\tilde{B}_i=B+\tilde{S}_i,
    \end{align}
\end{subequations}
where $\{\tilde{R}_i,\tilde{S}_i\}_{i=1}^\mathsf{N}$ are random variables that are independent with $A, B$. 
We assume that random noise variables $\{\tilde{R}_i,\tilde{S}_i\}_{i=1}^\mathsf{N}$ have zero means, and $\{\tilde{R}_i\}_{i=1}^\mathsf{N}$ and $\{\tilde{S}_i\}_{i=1}^\mathsf{N}$ are independent.
Node $i\in[\mathsf{N}]$ then outputs 
$
    \tilde{C}_i=\tilde{A}_i\tilde{B}_i.
$

We assume that up to $\mathsf{A}$ nodes can be adversarial, and up to $\mathsf{E}$ nodes can be erased. Specifically, for erased nodes $\mathcal{E}\subseteq[\mathsf{N}]$, $|\mathcal{E}|\le \mathsf{E}$, and adversarial nodes $\mathcal{A}\subseteq[\mathsf{N}]$, $|\mathcal{A}|\le \mathsf{A}$, 
a decoder receives an arbitrary element  from $\mathcal{Y}_{\mathcal{E},\mathcal{A}}$, where
$\mathcal{Y}_{\mathcal{E},\mathcal{A}}$ is the set of all vectors
$\begin{bmatrix}
    Y_1, Y_2, \cdots, Y_\mathsf{N}
\end{bmatrix}^T \in ( \mathbb{R} \cup \{\varepsilon\} )^\mathsf{N}$, satisfying
for $i \notin \mathcal{E} \cup \mathcal{A}$, $Y_i = \tilde{C}_i$; for $i\in \mathcal{E}$, $Y_i = \varepsilon$.
Here $\varepsilon$ represents an erasure symbol.
The decoder then applies a decoding function $d: ( \mathbb{R} \cup \{\varepsilon\} )^\mathsf{N}\rightarrow \mathbb{R}$ to estimate the product $A \cdot B$, i.e., given $\mathbf{Y} \in \mathcal{Y}_{\mathcal{E},\mathcal{A}}$, the decoder outputs $
    \tilde{C}=d(\mathbf{Y}).
$
We assume that the decoding function can be designed based on the knowledge of the joint distributions of $\{\tilde{R}_i,\tilde{S}_i\}_{i=1}^\mathsf{N}$ and parameters $\mathsf{N},\mathsf{E},\mathsf{A},\eta.$

A secure multiplication coding scheme $\mathcal{C}(\mathsf{N},\mathsf{A},\mathsf{E},\eta)$ specifies a joint distribution of $\{\tilde{R}_i,\tilde{S}_i\}_{i=1}^\mathsf{N}$ and a decoding functions $d:(\mathbb{R}\cup \{\varepsilon\})^{\mathsf{N}}\rightarrow \mathbb{R}$. We drop the dependence on the coding scheme $\mathcal{C}$ from parameters $(\mathsf{N},\mathsf{A},\mathsf{E},\eta)$ in this paper when the relationship is clear from the context. We measure the accuracy of a coding scheme $\mathcal{C}$ in terms of its maximum mean squared error over all erasure patterns and adversary actions.

\begin{Definition}[Mean square error] For a coding scheme $\mathcal{C}$ consisting of the joint distribution of $\{\tilde{R}_i,\tilde{S}_i\}_{i=1}^\mathsf{N}$ and a decoding function $d$, the mean square error ($\tt MSE$) is defined as 
\begin{align}
    {\tt MSE}(\mathcal{C}) = \sup_{\mathcal{E}\subseteq[\mathsf{N}],\mathcal{A}\subseteq[\mathsf{N}], |\mathcal{E}| \leq \mathsf{E}, |\mathcal{A}| \leq \mathsf{A}} {\tt MSE}_{\mathcal{E}, \mathcal{A}},
\end{align}
where
\begin{align}
    {\tt MSE}_{\mathcal{E}, \mathcal{A}} =  \mathbb{E}\left[ \sup_{\mathbf{Y} \in \mathcal{Y}_{\mathcal{E}, \mathcal{A}}} (d(\mathbf{Y})-AB)^2\right].
\end{align}

\end{Definition}

A coding scheme $\mathcal{C}(\mathsf{N},\mathsf{A},\mathsf{E},\eta)$ is said to satisfy $\mathsf{T}$-node $\epsilon$-DP if the data stored at any $\mathsf{T}$ nodes satisfies differential privacy with the parameter $\epsilon$ to the original input data. 
Mathematically:
\begin{Definition}[$\mathsf{T}$-node $\epsilon$-DP]\label{def::dp}
    A coding scheme $\mathcal{C}(\mathsf{N},\mathsf{A},\mathsf{E},\eta)$ with  random noise variables $\{\tilde{R}_i,\tilde{S}_i\}_{i=1}^N$ satisfies $\mathsf{T}$-node $\epsilon$-DP for $\epsilon>0$ if for any $A_0, B_0, A_1, B_1 \in \mathbb{R}$ satisfying $| A_0-A_1 | \le 1$, $| B_0-B_1 |\le 1$, we  have  
    \begin{align}
        \sup \left( \frac{\mathbb{P} \left( \mathbf{Y}_\mathcal{T}^{(0)} \in \mathcal{B}\right)}{\mathbb{P} \left( \mathbf{Y}_\mathcal{T}^{(1)} \in \mathcal{B}\right)}, \frac{\mathbb{P} \left( \mathbf{Z}_\mathcal{T}^{(0)} \in \mathcal{B}\right)}{\mathbb{P} \left( \mathbf{Z}_\mathcal{T}^{(1)} \in \mathcal{B}\right)} \right) \le e^{\epsilon},
    \end{align}
    for all subsets $\mathcal{T} \subseteq [\mathsf{N}]$ with $|\mathcal{T}|=\mathsf{T}$, and for all subsets $\mathcal{B}\subseteq \mathbb{R}^{1\times \mathsf{T}}$ in the Borel $\sigma$-field, where 
    $$
        \mathbf{Y}_\mathcal{T}^{(\ell)} \triangleq \begin{bmatrix}
            A_{\ell}+ \tilde{R}_{t_1} &  A_{\ell}+ \tilde{R}_{t_2} & \cdots & A_{\ell}+ \tilde{R}_{t_\mathsf{T}}
        \end{bmatrix},
        \qquad \mathbf{Z}_\mathcal{T}^{(\ell)} \triangleq \begin{bmatrix}
            B_{\ell}+ \tilde{S}_{t_1} &  B_{\ell}+ \tilde{S}_{t_2} & \cdots & B_{\ell}+ \tilde{S}_{t_\mathsf{T}}
        \end{bmatrix}
        $$
        with $\ell \in \{0,1\}$, $\mathcal{T}=\{t_1, t_2, \cdots, t_{\mathsf{T}}\}$.
\end{Definition}

{
The DP constraint ensures that, for any subset of $\mathsf{T}$ nodes, the joint distribution of $\{\tilde{A}_i\}_{i\in \mathcal{T}}$ satisfies $\epsilon$-DP. 
A similar constraint condition is imposed on the input $B$.
Hence, the privacy condition bounds the information that can be leaked about the private inputs $A$ and $B$ to the colluding nodes, ensuring that both inputs are protected under the DP constraint.
}

For fixed parameters $\mathsf{N},\mathsf{A},\mathsf{T},\mathsf{E}, \eta$, we are interested in studying the trade-off between the ${\tt MSE}$ and the $\mathsf{T}$-node DP parameter for coding schemes $\mathcal{C}(\mathsf{N},\mathsf{A},\mathsf{E},\eta).$

\subsection{Main Results}
    We denote $\left(\sigma^*(\epsilon) \right)^2$ as the smallest noise variance among all noise mechanisms achieving single-node $\epsilon$-DP. Specifically, for $\epsilon>0$, let $\mathcal{S}_{\epsilon}(\mathbb{P})$ denote the set of all real-valued random variables that satisfy $\epsilon$-DP, i.e., $X\in \mathcal{S}_{\epsilon}(\mathbb{P})$ if and only if,
\begin{align}
    \sup_{X',X'' \in \mathbb{R}, \mathcal{B} \subseteq \mathbb{R}, |X'-X''| \le 1} \frac{\mathbb{P}\left( X'+X \in \mathcal{B} \right)}{\mathbb{P}\left( X''+X \in \mathcal{B} \right)} \le e^{\epsilon}.
\end{align}
Let $L^2(\mathbb{P})$ denote the set of all real-valued random variables with finite variance. Then:
\begin{align}
    \left(\sigma^*(\epsilon) \right)^2 = \inf_{X\in \mathcal{S}_{\epsilon}(\mathbb{P}) \cap L^2(\mathbb{P})} \mathbb{E}\left[ (X-\mathbb{E}[X])^2 \right].
\end{align}

The optimal noise variance $\left(\sigma^*(\epsilon) \right)^2$ is characterized in \cite{geng2015optimal} as follows.
{\small\begin{align} \label{equ::optimal_noise}
    \left(\sigma^*(\epsilon) \right)^2=\frac{2^{2/3} e^{-2\epsilon/3}(1+e^{-2\epsilon/3})+e^{-\epsilon}}{(1-e^{-\epsilon})^2}.
\end{align}}

Our main result is the following theorem, which presents an achievable trade-off between the mean square error ${\tt MSE}(\mathcal{C})$ and the differential privacy parameter $\epsilon$ under the proposed coding scheme $\mathcal{C}$.


\begin{theorem}\label{thm::dp}
    For a positive integer $\mathsf{N}$ and non-negative integers $\mathsf{T},\mathsf{E}$ such that $\mathsf{N} \ge \mathsf{T}+\mathsf{E}+1$, there exists a secure multiplication scheme $\mathcal{C}$ that guarantees $\mathsf{T}$-node $\epsilon$-DP in presence of at most $\mathsf{E}$ erasures with the mean square error ${\tt MSE}(\mathcal{C})$ satisfying, for any $\gamma>0$,
    \begin{align} \label{equ::privacy_utility_tradeoff}
        {\tt MSE(\mathcal{C})} \le \frac{\eta^2}{\left(1+ {\tt SNR}^*\right)^2}+\gamma,
    \end{align}
    where ${\tt SNR^*} = \frac{\eta}{\left(\sigma^*(\epsilon) \right)^2}$.
\end{theorem}

See Section \ref{sec::dp_analysis} and \ref{sec::accuracy} for the proof.

\begin{theorem}\label{thm::adversary}
    For a positive integer $\mathsf{N}$ and non-negative integers $\mathsf{T},\mathsf{E},\mathsf{A}$ such that $\mathsf{N} \ge \mathsf{T}+\mathsf{E}+2\mathsf{A}+1$, there exists a secure multiplication scheme $\mathcal{C}$ along with an error decoding algorithm that guarantees $\mathsf{T}$-node $\epsilon$-DP in presence of at most $\mathsf{E}$ erasures and at most $\mathsf{A}$ adversaries, while achieving the privacy-utility tradeoff characterized in \eqref{equ::privacy_utility_tradeoff}.
\end{theorem}

See Section \ref{sec::error_correction} for the error decoding algorithm and the corresponding proof.

\begin{Remark}
    For the case where $\mathsf{N} \ge 2\mathsf{T}+\mathsf{E}+2\mathsf{A}+1$,${\tt MSE}(\mathcal{C})=0$ can be achieved by applying real-valued Shamir's secret sharing \cite{soleymani2021analog} for every $\epsilon>0$.
    For the special case where $\mathsf{E} = \mathsf{A} = 0,$ the results of Theorems \ref{thm::dp}, \ref{thm::adversary} were established in \cite{cadambe2023differentially}. The primary contribution of this paper addresses the general case where $\mathsf{E} \neq 0$ and $\mathsf{A} \neq 0$.
\end{Remark}

\begin{Remark}
    Previous work \cite{cadambe2023differentially}
    has established a lower bound $\frac{\eta^{2}}{(1+{\tt {SNR}}^{*})^2}$ of the ${\tt MSE}$ for all additive noise privacy mechanisms that satisfy $\mathsf{T}$-node $\epsilon$-DP with linear decoders, and has proven the tightness of this bound for its scheme. 
    A linear decoder, in this context, processes all outputs by replacing a subset of the coordinates affected by erasures or adversaries, then applies a linear operation to the remaining outputs. The coefficients of this linear operation depend solely on the remaining nodes.
    Our work achieves the same privacy-utility tradeoff as \cite{cadambe2023differentially} under the constraints of additive noises and linear decoders, and thus, the optimality of our approach is automatically implied by the results in \cite{cadambe2023differentially}.
\end{Remark}

\section{Proposed Coding Schemes} 
In this section, we introduce an $\mathsf{N}$-node secure multiplication coding scheme that achieves $\mathsf{T}$-node $\epsilon$-DP in the presence of at most $\mathsf{E}$ erased nodes and at most $\mathsf{A}$ adversarial nodes in Section \ref{sec::coding_schemes}. 
Here $\mathsf{T}, \mathsf{E}, \mathsf{A}$ are non-negative integers satisfying $\mathsf{N} \ge \mathsf{T}+\mathsf{E} + 2\mathsf{A} + 1 $. 
Section \ref{sec::dp_analysis} verifies the $\mathsf{T}$-node $\epsilon$-DP property. 
Upon receiving results from at least $\mathsf{N} - \mathsf{E}$ nodes, the decoder can resist up to $\mathsf{A}$ errors using the techniques outlined in Section \ref{sec::error_correction}.
In Section \ref{sec::accuracy}, we analyze the accuracy for estimating the product based on the decoded messages utilizing the algorithms introduced in Section~\ref{sec::error_correction}.

\subsection{Coding Schemes}\label{sec::coding_schemes}

Let $\{R_t, S_t\}_{t=1}^{\mathsf{T}}$ be statistically independent random variables with zero mean, and the choice of the distribution $\{R_t, S_t\}_{t=1}^{\mathsf{T}}$ will be specified in Section \ref{sec::dp_analysis} to guarantee $\mathsf{T}$-node $\epsilon$-DP for fixed DP parameters $\epsilon$.
We will then present a sequence of coding schemes indexed by positive integers $n$, that achieve the privacy-utility tradeoff described in Theorem \ref{thm::dp} as $n\rightarrow \infty$.
For the coding scheme with $\mathsf{T}> 1$ \footnote{The terms $\frac{1}{n}$ and $\frac{1}{n^{3/2}}$ in \eqref{equ::encoding_poly} can be replaced by other functions of $n$ as long as the constrains in (7) of \cite{cadambe2023differentially} are satisfied.}, let 
{
\begin{subequations}
    \label{equ::encoding_poly} 
    \begin{align}
    p_A(x)&=(A+R_1) + \frac{1}{n} \sum_{t=1}^\mathsf{T-1} R_{t+1}  x^t + \frac{1}{n^{3/2}} R_1 x^\mathsf{T}, \\
    p_B(x)&=(B+S_1) + \frac{1}{n}  \sum_{t=1}^\mathsf{T-1} S_{t+1} x^t + \frac{1}{n^{3/2}} S_1 x^\mathsf{T}.
    \end{align}    
\end{subequations}}
Select $\mathsf{N}$ distinct non-zero real numbers $\{x_i\}_{i=1}^\mathsf{N}$, and each node $i\in[\mathsf{N}]$ obtains noisy data as 
\begin{align}
    \tilde{A}_i = p_A(x_i), \quad \tilde{B}_i = p_B(x_i).
\end{align}
The coding scheme can be viewed as a $(\mathsf{N}, \mathsf{T}+1)$ real-valued RS code with messages $\{A+R_1, \frac{1}{n}R_2, \cdots, \frac{1}{n}R_{\mathsf{T}}, \frac{1}{n^{3/2}}R_1 \}$.

For the case with $\mathsf{T}=1$,  the received data of node $i\in \mathsf{N}$ can be represented as 
\begin{align}
    \tilde{A}_i = p_A(x_i) =(A+R_1) + \frac{1}{n^{3/2}} R_1 x_i, \quad 
    \tilde{B}_i = p_B(x_i) =(B+S_1) + \frac{1}{n^{3/2}} S_1 x_i.   
\end{align}

\begin{Remark}    \cite{cadambe2023differentially} proposed a coding scheme with $\mathsf{N}=\mathsf{T}+1$. 
    In this scheme, one designated node receives $A+R_1, B+S_1$ while the remaining $\mathsf{T}$ nodes receive $A+R_1 + \sum_{t=1}^\mathsf{T} O(\frac{1}{n})R_t + \frac{1}{n^{3/2}}R_1$ and $B+S_1 + \sum_{t=1}^\mathsf{T} O(\frac{1}{n})S_t + \frac{1}{n^{3/2}}S_1$. The output of the specially designated node is critical to the decoding process, making the design highly vulnerable, as the system fails to estimate the desired product if this node is erased or acts maliciously.
\end{Remark}

The intuition for the coding scheme design is as follows.
The noise added to the input $A$ can be interpreted as a superposition of three layers, distinguished by their magnitudes\footnote{Here we consider the setting with $\mathsf{T}>1$. For the case $\mathsf{T}=1$, only the first and the last layers of noise, i.e., $R_1$ and $\frac{1}{n^{3/2}}x_iR_1$, remain.}:
$R_1$ with magnitude $O(1)$, $\{\frac{1}{n} R_{t+1} x^t\}_{t=1}^{\mathsf{T}-1}$ with magnitude $O\left(\frac{1}{n} \right)$ and $\frac{1}{n^{3/2}} R_1 x^{\mathsf{T}}$ with magnitude $O\left(\frac{1}{n^{3/2}} \right)$. 
The first layer, $R_1$,  is carefully designed with an appropriate distribution and variance to ensure $\epsilon$-DP.
The third layer of noise, with magnitude $O\left(\frac{1}{n^{3/2}} \right)$ and correlated to the first layer, is to mitigate the negative impact of $R_1$ on accuracy. 
The second layer of noise of magnitude $O\left(\frac{1}{n}\right)$ prevents the collusion of up to $\mathsf{T}$ nodes from accessing the third layer.
This is achieved by letting $\lim_{n\rightarrow \infty} {\frac{1}{n^{3/2}}}/{\frac{1}{n}}=0$, effectively hiding the third layer. Simultaneously, $\mathsf{T}+1$ nodes can remove the second layer to improve estimation accuracy.

\subsection{Differential Privacy Analysis}\label{sec::dp_analysis}

Due to the symmetry of the proposed coding scheme, it suffices to demonstrate that the input $A$ satisfies $\mathsf{T}$-node $\epsilon$-DP and describe the design of additive noise $\{R_t\}_{t=1}^{\mathsf{T}}$.
The DP analysis for the input $B$ and the selection of $\{S_t\}_{t=1}^{\mathsf{T}}$ can be derived similarly.
To facilitate later analysis, we rewrite \eqref{equ::encoding_poly} based on the magnitude of each term as follows.
{
\begin{align}
    \tilde{A}_i &=\left(A+R_1\right)+\frac{1}{n}\begin{bmatrix}
        R_2  & \cdots & R_\mathsf{T}\
    \end{bmatrix} \mathbf{g}_i + \frac{1}{n^{3/2}} h_i R_1,
\end{align}}
where $\mathbf{g}_i = \begin{bmatrix}
    x_i & x_i^2 & \cdots & x_i^{\mathsf{T}-1}
\end{bmatrix}^T$ and $h_i = x_i^{\mathsf{T}}$.
Let $
\mathbf{G}= \begin{bmatrix}
    \mathbf{g}_1 & \mathbf{g}_2 & \cdots & \mathbf{g}_\mathsf{N} 
\end{bmatrix}^T 
$, $\mathbf{h}=\begin{bmatrix}
    h_1 & h_2 & \cdots & h_{\mathsf{N}}
\end{bmatrix}^T$, and then let the Vandermonde matrix $\mathbf{M}=\begin{bmatrix}
    \mathbf{1} & \mathbf{G} & \mathbf{h}
\end{bmatrix}$. 
Based on the property of Vandermonde matrix, every $(\mathsf{T}-1) \times (\mathsf{T}-1)$, $\mathsf{T} \times \mathsf{T}$ and $(\mathsf{T}+1)\times(\mathsf{T}+1)$ submatrix of $\mathbf{M}$ is guaranteed to be invertible.

We begin with the distributions of the independent noise variables $\{R_t\}_{t=1}^\mathsf{T}$.
For given DP parameter $\epsilon$, let $\sigma^2 = \left(\sigma^*(\epsilon) \right)^2+\gamma'$, where $\gamma'>0$.
For a fixed value $\sigma$, let $\epsilon^*$ be defined as,
{\begin{align}
    \epsilon^* = \inf_{Z, \mathbb{E}[Z^2] \ge \sigma^2} \sup_{\mathcal{B} \subseteq \mathbb{R}, A_0, A_1 \in \mathbb{R}, |A_0-A_1|\le 1} \ln \left( \frac{\mathbb{P}(A_0+ Z\in \mathcal{B})}{\mathbb{P}(A_1+ Z\in \mathcal{B})}\right),
\end{align}}
where $Z\in \mathbb{R}$ and $\mathbb{E}[Z]=0$.
Note that the noise variance $\mathbb{E}[Z^2]$ is strictly larger than $(\sigma^*(\epsilon))^2$. 
As $\sigma^*(\epsilon)$ strictly decreases with the DP parameter $\epsilon$ (according to \eqref{equ::optimal_noise}) and $\mathbb{E}[Z^2] > (\sigma^*(\epsilon))^2$, it follows that 
$\epsilon^* < \epsilon$.
For a DP parameter $\bar{\epsilon}$ with $\epsilon^* < \bar{\epsilon} < \epsilon$, there exists a random noise variable $Z^*$ such that $\mathbb{E}[(Z^*)^2]\le \sigma^2$satisfying,
{\begin{align}
    \sup_{\mathcal{B}\subseteq \mathbb{R}, -1<\lambda<1} \frac{\mathbb{P}(A+Z^* \in \mathcal{B})}{\mathbb{P}(A+Z^*+\lambda \in \mathcal{B})} \le e^{\bar{\epsilon}} \le e^{\epsilon}.
\end{align}}
Let the additive noise $R_1$ follow the same distribution as $Z^*$, and it follows that $A+R_1$ guarantees $\bar{\epsilon}$-DP. 
The noise variables $R_2, R_3, \cdots, R_\mathsf{T}$ are chosen as independent unit-variance Laplace random variables, each independent of $R_1$.

For the case with $\mathsf{T} \ge 2$, we assume that the first $\mathsf{T}$ nodes collude, i.e., the colluding node set is $\mathcal{T}=\{1,2 \cdots,\mathsf{T}\}$. For any other colluding set of $\mathsf{T}$ nodes, the argument we outline below will follow similarly due to the inherent symmetry in our coding scheme.
The colluding nodes receive:
$$
    \mathbf{Z}=(A +R_1)\mathbf{1}+ 
    \bar{\mathbf{G}}
    \begin{bmatrix}
        \frac{1}{n^{3/2}} R_1 & \frac{1}{n} R_2 & \cdots & \frac{1}{n}  R_\mathsf{T}
    \end{bmatrix}^T,
$$
where $\bar{\mathbf{G}}=\begin{bmatrix}
    h_1 & h_2 & \cdots & h_{\mathsf{T}} \\
    \mathbf{g}_1 & \mathbf{g}_2 & \cdots & \mathbf{g}_\mathsf{T}
\end{bmatrix}^T$.
Let $\mathbf{g'}_i^T$ with $i \in [\mathsf{T}]$ denote the $i$-th row of the matrix $\bar{\mathbf{G}}^{-1}$, and then $\mathbf{g'}_i^T\mathbf{1}$ represents the $i$-th element of the column vector $\bar{\mathbf{G}}^{-1}\mathbf{1}$.
We will now show that there is a full rank matrix $\mathbf{P}$ such that $\mathbf{Z}'=\mathbf{P} \mathbf{Z}$, where 
{
\begin{align}\label{equ::z'}
    \mathbf{Z}' = 
    \left[
        A + \left(1 + \frac{1}{ n^{3/2}\mathbf{g'}_1^T\mathbf{1}} \right) R_1 \quad   A + \frac{{1}+{n^{3/2}}\mathbf{g'}_1^T\mathbf{1}}{n\mathbf{g'}_2^T\mathbf{1}} R_2  \quad  \cdots \quad A + \frac{{1}+{n^{3/2}}\mathbf{g'}_1^T\mathbf{1}}{n\mathbf{g'}_\mathsf{T}^T\mathbf{1}} R_\mathsf{T}  
    \right]^T.
\end{align}}

As the matrix $\bar{\mathbf{G}}$ has a full rank of $\mathsf{T}$ according to the designed scheme, 
colluders can, through a one-to-one map of $\mathbf{Z}$ obtain:
$
    (A +R_1)\bar{\mathbf{G}}^{-1}\mathbf{1}+ 
    \begin{bmatrix}
        \frac{1}{n^{3/2}} R_1 & \frac{1}{n} R_2 & \cdots & \frac{1}{n}  R_\mathsf{T}
    \end{bmatrix}^T.
$

We now argue that $\mathbf{g'}_i^T\mathbf{1}\neq 0$.
Due to the fact that $\bar{\mathbf{G}}^{-1}\bar{\mathbf{G}}=\mathbf{I}$, we have that
$
\mathbf{g'}_1^T \begin{bmatrix}
        h_1 & h_2 & \cdots & h_\mathsf{T}
    \end{bmatrix}^T = 1
$
and
$
\mathbf{g'}_1^T
\begin{bmatrix}
    \mathbf{g}_1 & \mathbf{g}_2 & \cdots & \mathbf{g}_\mathsf{T}
\end{bmatrix}^T = \mathbf{0}^T.
$
The first equation shows that $\mathbf{g'}_1^T$ is not an all-zero row vector.
According to the coding scheme, the matrix $
\begin{bmatrix}
    1 & 1 & \cdots & 1 \\
    \mathbf{g}_1 & \mathbf{g}_2 & \cdots & \mathbf{g}_\mathsf{T}
\end{bmatrix}^T$ is full-rank, together with $
\mathbf{g'}_1^T
\begin{bmatrix}
    \mathbf{g}_1 & \mathbf{g}_2 & \cdots & \mathbf{g}_\mathsf{T}
\end{bmatrix}^T = \mathbf{0}^T
$,  $\mathbf{g'}_i^T\mathbf{1}$ cannot be zero.

We can then normalize the first component of the mapped $\mathbf{Z}$ and obtain $A + \left(1 + \frac{1}{ n^{3/2}\mathbf{g'}_1^T\mathbf{1}} \right) R_1$.
Next, we can use $A + \left(1 + \frac{1}{ n^{3/2}\mathbf{g'}_1^T\mathbf{1}} \right) R_1$ to remove $R_1$ terms \footnote{Note that we only consider the non-trivial case where $\mathbf{g'}_i^T\mathbf{1}\neq 0$ with $i\in \{2, \cdots, \mathsf{T}\}$. If $\mathbf{g'}_i^T\mathbf{1} = 0$, privacy is well-preserved as only noise remains.} in the other component of $\mathbf{Z}$. 
Hence we can derive $\mathbf{Z}'$ shown in \eqref{equ::z'}.

Let $\mathbf{Z}'= \begin{bmatrix}
    Z'_1 & Z'_2 & \cdots & Z'_\mathsf{T}
\end{bmatrix}^T$, and each $Z'_i$ with $i\in [\mathsf{T}]$ is expressed as a linear combination of $A$ and $R_i$. 
According to the post-processing property of differential privacy \cite{dwork2006differential} (performing arbitrary computations on the output of a DP mechanism does not increase the privacy loss), $\mathbf{Z}' = \mathbf{P} \mathbf{Z}$ inherits the DP guarantee of $\mathbf{Z}$, i.e., $\mathbf{Z}'$ remains DP with at least the same level of privacy as $\mathbf{Z}$.
To complete the proof, it therefore suffices to show that $\mathbf{Z}'$ is $\epsilon$-DP.

For $2 \le i\le \mathsf{T}$, the $i$-th term of $Z'_i$ is
$
A + \frac{{1}+{n^{3/2}}\mathbf{g'}_1^T\mathbf{1}}{n\mathbf{g'}_i^T\mathbf{1}} R_i
$,
where the second term represents a  Laplace random variable with variance $\left(\frac{{1}+{n^{3/2}}\mathbf{g'}_1^T\mathbf{1}}{n\mathbf{g'}_i^T\mathbf{1}} \right)^2$.
Since the added Laplace random noise variable with distribution $\text{Lap}(\frac{1}{\epsilon})$ ensures ${\epsilon}$-DP \cite{dwork2006calibrating}, $Z'_i$ serves as a privacy mechanism achieving $\frac{n\mathbf{g'}_i^T\mathbf{1}}{{1}+{n^{3/2}}\mathbf{g'}_1^T\mathbf{1}} \sqrt{2}$-DP as  $R_2, R_3, \cdots, R_\mathsf{T}$ are independent unit-variance Laplace random variables.

For $i=1$, we have that 
{
\begin{align}
    &\sup_{\mathcal{B}\subseteq \mathbb{R}, -1<\lambda<1} \frac{\mathbb{P}\left(A + \left(1 + \frac{1}{ n^{3/2}\mathbf{g'}_1^T\mathbf{1}} \right) R_1 \in \mathcal{B}\right)}{\mathbb{P}\left(A + \left(1 + \frac{1}{ n^{3/2}\mathbf{g'}_1^T\mathbf{1}} \right) R_1+\lambda \in \mathcal{B}\right)} \nonumber \\
    =&
    \sup_{\mathcal{B}\subseteq \mathbb{R}, -\frac{1}{1+\frac{1}{\mathbf{g'}_1^T\mathbf{1}}\frac{1}{n^{3/2}}}<\lambda<\frac{1}{1+\frac{1}{\mathbf{g'}_1^T\mathbf{1}}\frac{1}{n^{3/2}}}} \frac{\mathbb{P}(A+R_1 \in \mathcal{B})}{\mathbb{P}(A+R_1+\lambda \in \mathcal{B})} \nonumber \\
    \overset{(a)}{\le} & e^{\bar{\epsilon}}+\gamma',
\end{align}}
where $(a)$ holds as $\lim_{n\rightarrow \infty} \frac{1}{1+\frac{1}{n^{3/2}\mathbf{g'}_1^T\mathbf{1}}}$, for any $\gamma'>0$. Hence $Z'_1$  achieves $\bar{\epsilon}$-DP.

Since $R_1, R_2, ...,R_\mathsf{T}$ are independent, $\mathbf{Z}'$ achieves $(\bar{\epsilon}+ \sqrt{2}\sum_{i=2}^\mathsf{T} \frac{n\mathbf{g'}_i^T\mathbf{1}}{{1}+{n^{3/2}}\mathbf{g'}_1^T\mathbf{1}})$-DP by the composition theorem \cite{dwork2006differential}. As we have $\lim_{n\rightarrow \infty} \frac{n}{n^{3/2}} = 0$, the DP parameter converges to $\bar{\epsilon}$ as $n\rightarrow \infty$.
The coding scheme hence ensures $\epsilon$-DP.

For the case with $\mathsf{T}=1$, we have that 
\begin{align}
    &\sup_{\mathcal{B}\subseteq \mathbb{R}, -1<\lambda<1} \frac{\mathbb{P}\left(A + \left(1 + \frac{1}{ n^{3/2}} h_i \right) R_1 \in \mathcal{B}\right)}{\mathbb{P}\left(A + \left(1 + \frac{1}{ n^{3/2}} h_i \right) R_1+\lambda \in \mathcal{B}\right)} \nonumber \\
    =&
    \sup_{\mathcal{B}\subseteq \mathbb{R}, -\frac{1}{1 + \frac{1}{ n^{3/2}} h_i}<\lambda<\frac{1}{1 + \frac{1}{ n^{3/2}} h_i}} \frac{\mathbb{P}(A+R_1 \in \mathcal{B})}{\mathbb{P}(A+R_1+\lambda \in \mathcal{B})} \nonumber \\
    \overset{(a)}{\le} & e^{\bar{\epsilon}}+\gamma' \le  e^{\epsilon},
\end{align}
where $(a)$ holds as $\lim_{n\rightarrow \infty} \frac{1}{1 + \frac{1}{ n^{3/2}} h_i} =1$ for any $\gamma'>0$. The coding scheme still guarantees $\epsilon$-DP.
Hence, we have proved the proposed scheme satisfies $\mathsf{T}$-node $\epsilon$-DP.

\subsection{Accuracy Analysis} \label{sec::accuracy}
{Recall that node $i\in [\mathsf{N}]$ computes $\tilde{C}_i = p_A(x_i) p_B(x_i)$, where polynomials $p_A(x)$ and $p_B(x)$ are specified in  \eqref{equ::encoding_poly}.
Based on the definitions of the polynomials $p_A(x)$ and $p_B(x)$ in \eqref{equ::encoding_poly}, the readers can verify that each $\tilde{C}_i$ can be rewritten as: 
\begin{align}\label{equ::polynomial_C}
    \tilde{C}_{i} = P(x_{i}) + \frac{r_{i}}{n^2} + \frac{t_{i}}{n^{\frac{5}{2}}}+\frac{u_{i}}{n^{3}},
\end{align}
where $\{r_{i}\}_{i\in [\mathsf{N}]}$, $\{t_{i}\}_{i\in [\mathsf{N}]}$, and $\{u_{i}\}_{i\in [\mathsf{N}]}$ are determined by the polynomial multiplication and are independent of $n$,
$P(x)=\sum_{j=0}^{\mathsf{T}}  m_j x^j$ is a degree-$\mathsf{T}$ polynomial with coefficients:
\begin{align}\label{equ::m}
    &m_0 =  \left(A+ R_1 \right)\left(B+ S_1 \right), m_1 = \frac{1}{n} \left(S_2(A+R_1)+R_2(B+S_1) \right), \cdots, m_{\mathsf{T}-1}= \frac{1}{n} (S_{\mathsf{T}}(A+R_1)+R_{\mathsf{T}}(B+S_1) ), \nonumber \\
    &m_{\mathsf{T}}=\frac{1}{n^{3/2}} (S_1(A+R_1)+R_1(B+S_1) ).
\end{align}
In Section \ref{sec::error_correction} and Appendix \ref{apd::distortion}, we will show that based on the receiving results from $\mathsf{N} - \mathsf{E}$ non-erased nodes, the decoder can effectively determine $\mathsf{T}+1$ accurate and available computation outputs.
Without loss of generality, we conduct an accuracy analysis assuming that node $i\in [\mathsf{T}+1]$ is not erased or adversarial in this subsection.
The message vector $\tilde{\mathbf{C}}=\begin{bmatrix}
    \tilde{C}_1 & \tilde{C}_2 & \cdots & \tilde{C}_{\mathsf{T}+1}
\end{bmatrix}^T$ can be rewritten as
\begin{align}
    \tilde{\mathbf{C}} = \mathbf{V} \begin{bmatrix}
         m_0 & m_1 & \cdots & m_\mathsf{T}
    \end{bmatrix}^T
    + O\left(\frac{1}{n^2}\right)\mathbf{1},
\end{align}
where $\mathbf{V}$ is a $(\mathsf{T}+1)\times (\mathsf{T}+1)$ Vandermonde matrix with distinct evaluation points $\{x_i\}_{i=1}^{\mathsf{T}+1}$, and coefficients $\{m_j\}_{j=0}^\mathsf{T}$ are specified in \eqref{equ::m}. 
As the Vandermonde matrix is invertible, the decoder can obtain $\{\tilde{m}_j \}_{j=0}^{\mathsf{T}}$ which are the approximation of $\{{m}_j \}_{j=0}^{\mathsf{T}}$ by calculating $\mathbf{V}^{-1} \tilde{\mathbf{C}}$.

Utilizing $\tilde{m}_0$ and $\tilde{m}_{\mathsf{T}}$, the decoder could get,
\begin{subequations}\label{equ::C1C2}
    \begin{align}
        & \bar{C}_1= \tilde{m}_0 =\left(A+ R_1 \right)\left(B+ S_1 \right) + O\left(\frac{1}{n^2}\right),\\
        &  \bar{C}_2= \tilde{m}_0 +\tilde{m}_{\mathsf{T}}=\left(A+ \left(1+\frac{1}{n^{3/2}}\right) R_1 \right)\left(B+ \left(1+\frac{1}{n^{3/2}}\right) S_1 \right)  +O\left(\frac{1}{n^2}\right).
    \end{align}
\end{subequations}
}

The following useful lemma is a well-known result from linear mean square estimation theory\cite{poor2013introduction}.
\begin{lemma}\label{lemma::lmse}
    Let $X$ be a random variable with $\mathbb{E}[X]=0$ and $\mathbb{E}[X^2]=\lambda^2$. Let $\{N_i\}_{i=1}^m$ be random noise variables independent of $X$, and
    $\tilde{\mathbf{X}}=\begin{bmatrix}
        \nu_1 X+ N_1 & \cdots & \nu_m X+ N_m
    \end{bmatrix}^T$, where $\nu_i \in \mathbb{R}$.
    Then 
    $\inf_{\mathbf{w}\in\mathbb{R}^{m}} \mathbb{E}[| \mathbf{w}^T \tilde{\mathbf{X}} -X|^2]=\frac{\lambda^2}{1+{\tt SNR}_a},$
    and  
    $
        {\tt SNR}_a=\frac{\det (\mathbf{K}_1)}{\det(\mathbf{K}_2)}-1
    $, where $\mathbf{K}_1$ denotes the covariance matrix of the noisy observation $\tilde{\mathbf{X}}$, and $\mathbf{K}_2$ denotes the covariance matrix of the noise $\{N_i\}_{i=1}^m$.
    Furthermore, there exists a vector $\mathbf{w}^*\in \mathbb{R}^{m}$ such that the linear mean square error achieves  $\frac{\lambda^2}{1+{\tt SNR}_a}$.
    Also, the vector $\mathbf{w}^*$ satisfies that for any random variables $X'$ with  $\mathbb{E}[X']=0$ and $\mathbb{E}[X'^2] \le \lambda^2$, $\mathbb{E}[| \mathbf{w}^{*T} \tilde{\mathbf{X}}'-X'|^2] \le \frac{\lambda^2}{1+{\tt SNR}_a}$, where $\tilde{\mathbf{X}}'=\begin{bmatrix}
        \nu_1 X' + N_1 & \cdots & \nu_m X'+ N_m
    \end{bmatrix}^T$.
\end{lemma}

From Lemma \ref{lemma::lmse}, we can infer that there exists a linear decoder that obtains $\tilde{C}$ from $\bar{C}_1, \bar{C}_2$  with a mean squared error not exceeding $\frac{\eta^2}{1+{\tt SNR}_a}$, where ${\tt SNR}_a$ is defined in \eqref{equ::snr_accuracy}. 
\begin{align}\label{equ::snr_accuracy}
    {\tt SNR}_a &= \frac
    {\left | \begin{matrix}
    \eta^2+2\eta \sigma^2  + \sigma^4+O\left(\frac{1}{n^4}\right)
    &\eta^2+2\eta \sigma^2 \left(1+\frac{1}{n^{3/2}} \right) + \sigma^4\left(1+\frac{1}{n^{3/2}} \right)^2+O\left(\frac{1}{n^4}\right)\\
    \eta^2+2\eta \sigma^2 \left(1+\frac{1}{n^{3/2}} \right) + \sigma^4\left(1+\frac{1}{n^{3/2}} \right)^2+O\left(\frac{1}{n^4}\right) &  \eta^2+2\eta\sigma^2 \left(1+\frac{1}{n^{3/2}} \right)^2 + \sigma^4 \left(1+\frac{1}{n^{3/2}} \right)^4 +O\left(\frac{1}{n^4}\right)\\
    \end{matrix} \right | }
    {\left | \begin{matrix}
    2\eta \sigma^2  + \sigma^4+O\left(\frac{1}{n^4}\right)
    &2\eta \sigma^2 \left(1+\frac{1}{n^{3/2}} \right) + \sigma^4\left(1+\frac{1}{n^{3/2}} \right)^2+O\left(\frac{1}{n^4}\right)\\
    2\eta \sigma^2 \left(1+\frac{1}{n^{3/2}} \right) + \sigma^4\left(1+\frac{1}{n^{3/2}} \right)^2+O\left(\frac{1}{n^4}\right)&  2\eta\sigma^2 \left(1+\frac{1}{n^{3/2}} \right)^2 + \sigma^4 \left(1+\frac{1}{n^{3/2}} \right)^4 +O\left(\frac{1}{n^4}\right)\\
    \end{matrix} \right | } -1 \nonumber\\
    &\overset{(a)}{=}\frac{\eta^2+2\eta \sigma^2+2\frac{1}{n^{3/2}} \eta \sigma + O(\frac{1}{n^2})}
    {\sigma^4\left(1+\frac{1}{n^{3/2}} \right)^2+O(\frac{1}{n^2})} \nonumber \\
    &\overset{(b)}{=}\frac{\eta^2}{\sigma^4}+\frac{2\eta}{\sigma^2} + O\left(\frac{1}{n}\right).
\end{align}
where $(a)$ holds by omitting $o\left(\frac{1}{n^2}\right)$ term, $(b)$ holds as $\lim_{n\rightarrow \infty} \frac{1}{n^{3/2}} = 0$ and $\lim_{n\rightarrow \infty} \frac{1}{n^2} = 0$.
As $O\left(\frac{1}{n}\right)$ means the term tends to 0 with sufficiently large $n$, the lower bound of ${\tt SNR}_a$ is derived, i.e.,
{
\begin{align}\label{equ::snra_bound}
    {\tt SNR}_a \ge \frac{\eta^2}{\sigma^4}+\frac{2\eta}{\sigma^2}-\gamma',
\end{align}}
for any $\gamma'>0$ by selecting sufficiently large $n$.
The same result can be easily derived when $\mathsf{T}=1$ by utilizing $m_0$ and $m_1$.

{
Specifically, the linear minimum mean squared error (MMSE) \cite{kay1993fundamentals} estimate of $C$ based on $\tilde{C}_1$ and $\tilde{C}_2$ can be expressed as
 $\tilde{C}=d_1 \bar{C}_1 + d_2 \bar{C}_2 $, where the optimal $d_1$ and $d_2$ are obtained by solving
 \begin{align}\label{equ::optimal_estimater}
     \begin{bmatrix}
        d_1 \\
        d_2
    \end{bmatrix} = 
    \begin{bmatrix}
    \eta^2+2\eta \sigma^2  + \sigma^4
    &\eta^2+2\eta \sigma^2 \left(1+\frac{1}{n^{3/2}} \right) + \sigma^4\left(1+\frac{1}{n^{3/2}} \right)^2\\
    \eta^2+2\eta \sigma^2 \left(1+\frac{1}{n^{3/2}} \right) + \sigma^4\left(1+\frac{1}{n^{3/2}} \right)^2& \eta^2+2\eta\sigma^2 \left(1+\frac{1}{n^{3/2}} \right)^2 + \sigma^4 \left(1+\frac{1}{n^{3/2}} \right)^4\\
    \end{bmatrix}^{-1}
    \begin{bmatrix}
        \eta^2 \\
        \eta^2
    \end{bmatrix}
 \end{align}
}

Combining Lemma \ref{lemma::lmse} and \eqref{equ::snra_bound}, the achievable result in Theorem \ref{thm::dp} is proved by substituting $\sigma^2 = \left(\sigma^*(\epsilon) \right)^2+\gamma''$ with sufficient small $\gamma''$ as shown in Section \ref{sec::dp_analysis}.

\subsection{Error Correction Methods}\label{sec::error_correction}
{
Assuming that at most $\mathsf{E}$ erasures can occur, the decoder is guaranteed to receive at least $\mathsf{N}-\mathsf{E} \ge \mathsf{T}+2\mathsf{A}+1$ computation results from surviving nodes (non-erasures). 
We arbitrarily select $\mathsf{T}+2\mathsf{A}+1$ available symbols received from surviving nodes, and let $\mathcal{S}=\{s_{1}, s_2, \cdots, s_{\mathsf{T}+2\mathsf{A}+1}\}$ denote the set of corresponding indices, where $|\mathcal{S}| = \mathsf{T}+2\mathsf{A}+1$  and $\{s_i\}_{i=1}^{\mathsf{T}+2\mathsf{A}+1}$ represent the selected indices.
The decoder receives the vector $[Y_{s_1}, Y_{s_2}, \cdots, Y_{s_{\mathsf{T}+2\mathsf{A}+1}}]^T \in \mathbb{R}^{\mathsf{T}+2\mathsf{A}+1}$, with at most $\mathsf{A}$ of them adversarially corrupted.
Let $\mathcal{A} \subseteq \mathcal{S}$ be the set of indices corresponding to adversarial nodes, defined as $\mathcal{A}=\{i|Y_{s_i} \neq \tilde{C}_{s_i}, i \in[\mathsf{T}+2\mathsf{A}+1]\}$ with $|\mathcal{A}|\le \mathsf{A}$. 
In this subsection, we consider the sequence of coding schemes indexed by $n$ with the evaluation points fixed, and aim to show that these $\mathsf{T}+2\mathsf{A}+1$  received values are sufficient to recover the desired multiplication result at the $\tt{MSE}$ bound characterized in Theorem \ref{thm::dp} while tolerating up to $\mathsf{A}$ adversarial errors.

Note that the computation result from node $i\in \mathsf{N}$, $\tilde{C}_i$ can be expressed as $\tilde{C}_i = p_A(x_i) p_B(x_i)$, i.e., the evaluation of a degree-$2\mathsf{T}$ polynomial at the point $x_i$.
Based on the classical coding theory \cite{huffman2010fundamentals}, correcting $\mathsf{A}$ adversarial errors using a RS code of degree $\mathsf{2T}$ requires at least $2\mathsf{T}+2\mathsf{A}+1$ evaluations, while under the system parameter assumption $\mathsf{N} \ge \mathsf{T}+\mathsf{E}+2\mathsf{A}+1$, reliable recovery of the multiplication result must be guaranteed using only $\mathsf{T}+2\mathsf{A}+1$ evaluations.
To satisfy this constraint, we exploit the observation that the product polynomial $p_A(x)p_B(x)$ contains $\mathsf{T}$ higher-order coefficients whose magnitudes are $O\left(\frac{1}{n^2}\right)$.
Informally speaking, we can neglect these $\mathsf{T}$ terms, and approximately treat $p_A(x)p_B(x)$ as a degree-$\mathsf{T}$ polynomial rather than a polynomial of degree $2\mathsf{T}$, as shown in \eqref{equ::polynomial_C}.\footnote{{This approximation is also justified in the accuracy analysis presented in Section \ref{sec::accuracy}, where the $O\left(\frac{1}{n^2}\right)$ terms are shown to have a negligible effect on overall accuracy.}} 
Hence $\{\tilde{C}_{s_i}\}_{i=1}^{\mathsf{T}+2\mathsf{A}+1}$ can be approximately viewed as a $(\mathsf{T}+2\mathsf{A}+1, \mathsf{T}+1)$ RS code with respect to encoding coefficients $\{m_j\}_{j=0}^{\mathsf{T}}$ specified in \eqref{equ::m} and the corresponding degree-$\mathsf{T}$ encoding polynomial $P(x)=\sum_{j=0}^\mathsf{T} m_j x^j$.

We adapt the Berlekamp–Welch algorithm \cite{welch1986error} to handle real-valued outcomes for decoding (see Algorithm \ref{alg::error_locating}).
We first briefly recall the Berlekamp–Welch decoding algorithm applied to the encoding polynomial $P(x)$.
Let
\begin{align}\label{equ::polynomial_E}
    {E}(x) =  {e}_0 +  {e}_1x + \cdots + {e}_{\mathsf{A}-1}x^{\mathsf{A}-1} + x^{\mathsf{A}}
\end{align}
be a monic error locator polynomial of degree $\mathsf{A}$, where ${E}(x_{s_i})=0$ if node $s_i$ is adversarial.
Let 
\begin{align}\label{equ::polynomial_Q}
    {Q}(x) =  {q}_0 +  {q}_1x+ \cdots + {q}_{\mathsf{T}+\mathsf{A}} x^{\mathsf{T}+\mathsf{A}}
\end{align}
be  the product of the error locator polynomial $ {E}(x)$ and encoding polynomial $P(x)$, and the degree of $ {Q}(x)$ is $\mathsf{T}+\mathsf{A}$.
Note the $\mathsf{T}+2\mathsf{A}+1$ coefficients $\{{e}_j\}_{j=0}^{\mathsf{A}-1}$ and $\{q_j\}_{j=0}^{\mathsf{T}+\mathsf{A}}$ are initially unknown and determined via the $\mathsf{T}+2\mathsf{A}+1$ observations in the decoding process. Once both $E(x)$ and $Q(x)$ are recovered, the encoding polynomial $P(x)$ can be derived in the classical setting by solving $P(x)=Q(x)/E(x)$.

{
In our setting, we modify the Berlekamp-Welch algorithm (see Algorithm \ref{alg::error_locating} for details) as follows to account for approximation. The decoder collects values $\{Y_{s_i}\}_{i=1}^{\mathsf{T}+2\mathsf{A}+1}$ at the corresponding evaluation points $\{x_{s_i}\}_{i=1}^{\mathsf{T}+2\mathsf{A}+1}$ and solves the linear system:
\begin{align} \label{equ::actual_linear_system_general}
    \tilde{\mathbf{D}} \mathbf{\tilde{b}} = \mathbf{\tilde{c}},
\end{align}
where
$\tilde{\mathbf{D}} = \begin{bmatrix}
        Y_{s_1}  & \cdots & x_{s_1}^{\mathsf{A-1}}Y_{s_1} & -1 &  \cdots & -x_{s_2}^{\mathsf{T}+\mathsf{A}}\\
        Y_{s_2}  & \cdots & x_{s_2}^{\mathsf{A-1}}Y_{s_2} & -1 &  \cdots & -x_{s_2}^{\mathsf{T}+\mathsf{A}}\\
        \vdots  & \ddots & \vdots  & \vdots & \ddots & \vdots\\
        Y_{s_{\mathsf{T}+2\mathsf{A}+1}}  & \cdots & x_{s_{\mathsf{T}+2\mathsf{A}+1}}^{\mathsf{A-1}}Y_{s_{\mathsf{T}+2\mathsf{A}+1}} & -1  & \cdots & -x_{s_{\mathsf{T}+2\mathsf{A}+1}}^{\mathsf{T}+\mathsf{A}}
    \end{bmatrix}_{(\mathsf{T}+2\mathsf{A}+1)\times (\mathsf{T}+2\mathsf{A}+1)}$, 
$\mathbf{\tilde{c}}=\begin{bmatrix}
    -x_{s_1}^\mathsf{A} Y_{s_1} \\
    \vdots \\
    -x_{s_{\mathsf{T}+2\mathsf{A}+1}}^\mathsf{A} Y_{s_{\mathsf{T}+2\mathsf{A}+1}}
\end{bmatrix}_{\mathsf{T}+2\mathsf{A}+1}$.
Solving this system yields the estimated coefficient vector $\mathbf{\tilde{b}} = 
\begin{bmatrix}
    \tilde{e}_0 &
    \cdots &
    \tilde{e}_{\mathsf{A}-1} &
    \tilde{q}_0 &
    \cdots &
    \tilde{q}_{\mathsf{T}+\mathsf{A}}
\end{bmatrix}^T$, which is the approximation of coefficients $\{{e}_j\}_{j=0}^{\mathsf{A}-1}$ and $\{{q}_j\}_{j=0}^{\mathsf{T}+\mathsf{A}}$.
Let $\tilde{E}(x)$ denote the polynomial 
determined by the coefficients $\{\tilde{e}_j\}_{j=0}^{\mathsf{A}-1}$ and $\tilde{e}_\mathsf{A}=1$, i.e., 
\begin{align}
    \tilde{E}(x) =  \tilde{e}_0 +  \tilde{e}_1 x + \cdots + \tilde{e}_{\mathsf{A}-1}x^{\mathsf{A}-1} + x^{\mathsf{A}}.
\end{align}

}

The polynomial $\tilde{E}(x)$ is constructed from the estimated coefficients $\{\tilde{e}_j\}_{j=0}^{\mathsf{A}-1}$.
The decoder then ranks the values of $|\tilde{E}(x_{s_i})|$ in descending order and selects the $\mathsf{T}+1$ evaluation points corresponding to the $\mathsf{T}+1$ largest values. Utilizing the selected points, the decoder applies the method detailed in Section~\ref{sec::accuracy} to compute the estimated result $\tilde{C}$.
The details are illustrated in Algorithm \ref{alg::error_locating}.

The following proposition illustrates the type of adversaries that can be excluded by Algorithm \ref{alg::error_locating}. 
{
\begin{Proposition}\label{prop::main}
      For an adversarial node $s_i \in \mathcal{A}$ with distortion $\Delta_{s_i} = |P(x_{s_i}) - Y_{s_i}| \ge n^{-1.6}$, the probability that $s_i$ is selected in the top $t+1$ nodes (\ref{line:top_selection} of Algorithm \ref{alg::error_locating}) tends to 0 as $n \to \infty$.
\end{Proposition}}

\begin{proof}
    See Appendix \ref{apd::distortion}.
\end{proof}

If an adversarial node $s_i \in \mathcal{A}$ does not satisfy the condition that $\Delta_{s_i} \ge n^{-1.6}$, it may avoid detection and the corresponding error may not be excluded by Algorithm \ref{alg::error_locating}. However, such a node contributes negligibly to the approximation error, enabling the desired accuracy as $n\to\infty$.

We then prove Theorem \ref{thm::adversary}, which shows the effectiveness of the error decoding algorithm.

\begin{proof}

For each adversarial node $s_i \in \mathcal{A}$, we distinguish two cases based on the distortion magnitude $\Delta_{s_i}$, depending on whether $\Delta_{s_i} \ge n^{-1.6}$ or $\Delta_{s_i} < n^{-1.6}$. We analyze these two scenarios when $n$ tends to infinity.
\begin{enumerate}
    \item If $\Delta_{s_i} \ge n^{-1.6}$, then by Proposition~\ref{prop::main}, such an adversary is excluded by Algorithm~\ref{alg::error_locating} with probability approaching 1 as $n \to \infty$.
    As detailed in Appendix~\ref{apd::distortion}, the case where an adversary may evade detection arises from rare extreme realizations of the random variables $\{R_i, S_i \}_{i=1}^\mathsf{T}$.
    Since these variables are independent of the multiplicands $A, B$, and both $A, B$ have bounded second moments, such exceptional cases,  which occur with negligible probability, do not affect the upper bound on the ${\tt MSE}$.
    \item If $\Delta_{s_i} < n^{-1.6}$, the adversary may evade detection by Algorithm~\ref{alg::error_locating}. 
    In this case, the decoder could obtain the following quantities, instead of those in \eqref{equ::C1C2},
    \begin{subequations}
    \begin{align}
        & \bar{C}_1'= \tilde{m}_0 =\left(A+ R_1 \right)\left(B+ S_1 \right) + O\left(\frac{1}{n^{1.6}}\right),\\
        &  \bar{C}_2'= \tilde{m}_0 +\tilde{m}_{\mathsf{T}}=\left(A+ \left(1+\frac{1}{n^{3/2}}\right) R_1 \right)\left(B+ \left(1+\frac{1}{n^{3/2}}\right) S_1 \right)  +O\left(\frac{1}{n^{1.6}}\right).
    \end{align}
\end{subequations}
    Following the similar ${\tt SNR}$ analysis in \eqref{equ::snr_accuracy}, such a distortion level is asymptotically negligible and does not significantly affect the achievable ${\tt SNR_a}$.
\end{enumerate}
Hence for any $\xi>0$, by selecting sufficiently large $n$, it follows that
\begin{align}
    {\tt SNR}_a \ge \frac{\eta^2}{\sigma^4}+\frac{2\eta}{\sigma^2}-\xi.
\end{align}
Together with Lemma \ref{lemma::lmse}, the privacy-utility tradeoff in \eqref{equ::privacy_utility_tradeoff} can be achieved when $n \to \infty$, and Theorem \ref{thm::adversary} is proved.

\end{proof}
}

\begin{Remark}
The classical Berlekamp-Welch algorithm operates over a finite field, and the original message can, in principle, be recovered either via polynomial division, $P(x) = Q(x)/E(x)$, or by interpolating over non-adversarial evaluation points.
In this work, we focus on decoding in the real domain. The presence of perturbations on the order of $O(1/n^2)$ introduces challenges that the approximate encoding polynomial cannot be reliably recovered through direct division, as the result may no longer be a polynomial. Consequently, the decoding strategy must be carefully designed. 
The error correction codes over the real field are explored in \cite{roth2020analog}, where errors exceeding a certain threshold are detected and corrected and noise with small amplitude can be tolerated. 
In contrast to classical error correction, the goal of this work is not to identify all adversary nodes but to ensure that the privacy-utility tradeoff in Theorem \ref{thm::dp} will not be compromised in the presence of some adversaries. 
Similar to \cite{roth2020analog}, we exclude adversarial nodes that introduce significant distortion (i.e., those for which $\Delta_{s_i} \ge n^{-2+\alpha}$ for some $\alpha > 0$) while tolerating small noise. 
To this end, we adopt a method that selects $t+1$ evaluation points that maximize $|\tilde{E}(x)|$, and we provide a theoretical justification.
\end{Remark}

{

\begin{Remark}
  Unlike the approach in \cite{soleymani2022approxifer}, which detects adversarial nodes by selecting the $\mathsf{A}$ evaluation points that minimize $|\tilde{E}(x_{s_i})|$, we instead select the $\mathsf{T}+1$ evaluation points with the largest values of $|\tilde{E}(x_{s_i})|$ and designate them as non-adversarial. 
  The proposed approach is theoretically justified within the considered secure multiplication setting.
\end{Remark}
}


\begin{algorithm}[tbp]\label{alg::error_locating}
    \SetKwInOut{Input}{Input}
    \SetKwInOut{Output}{Output}
    \SetKwInOut{Require}{Require}
    \caption{Error Decoding Algorithm}
    \SetAlgoLined
    {\Require{number of nodes $\mathsf{N}$, number of colluding nodes $\mathsf{T}$, maximum number of erased nodes $\mathsf{E}$, maximum number of adversarial nodes $\mathsf{A}$, evaluation points $\{x_i\}_{i=1}^\mathsf{N}$, $\eta$
    }
    \Input{a random selection of $\mathsf{T}+2\mathsf{A}+1$ non-erased received values, denoted by $\{Y_{s_i}\}_{i=1}^{\mathsf{T}+2\mathsf{A}+1}$.} 
    \Output{the estimated multiplication result $\tilde{C}$.}
    Formulate $\tilde{\mathbf{D}}$ and $\tilde{\mathbf{c}}$ in \eqref{equ::actual_linear_system_general} based on $\{Y_{s_i}\}_{i=1}^{\mathsf{T}+2\mathsf{A}+1}$ and $\{x_{s_i}\}_{i=1}^{\mathsf{T}+2\mathsf{A}+1}$. \\
    Solve the linear system $\tilde{\mathbf{D}} \mathbf{\tilde{b}} = \mathbf{\tilde{c}}$ to get $\mathbf{\tilde{b}}$.\\
    Extract $\{\tilde{e}_j\}_{j=0}^{\mathsf{A}-1}$ from $\mathbf{\tilde{b}}$, and set  $\tilde{e}_\mathsf{A}=1$, to define the polynomial $\tilde{E}(x)=\sum_{j=0}^\mathsf{A} \tilde{e}_j x^j$. \\
    \For{$i = 1, ..., \mathsf{T}+2\mathsf{A}+1$}
    {Calculate $\tilde{E}(x_{s_i}) = \tilde{e}_0 +  \tilde{e}_1x_{s_i} + \cdots + \tilde{e}_{\mathsf{A}-1}x_{s_i}^{\mathsf{A}-1} + x_{s_i}^{\mathsf{A}}$.}
    Sort $\left|\tilde{E}(x_{s_i}) \right|$ decreasingly, i.e., $\left|\tilde{E}(x_{a_1})\right| \ge \left|\tilde{E}(x_{a_2})\right| \ge \cdots \ge \left|\tilde{E}(x_{a_{\mathsf{T}+2\mathsf{A}+1}})\right|$.\\
    Obtain the estimated multiplication result $\tilde{C}$ using the method described in Section \ref{sec::accuracy} based on messages $\{Y_i\}_{i\in\{a_1, \cdots, a_{\mathsf{T}+1}\} }$ 
    \label{line:top_selection}\\
    }
\end{algorithm}

\begin{figure}[H]
    \centering
    \begin{minipage}[t]{0.48\textwidth}
        \centering
        \includegraphics[scale=0.47]{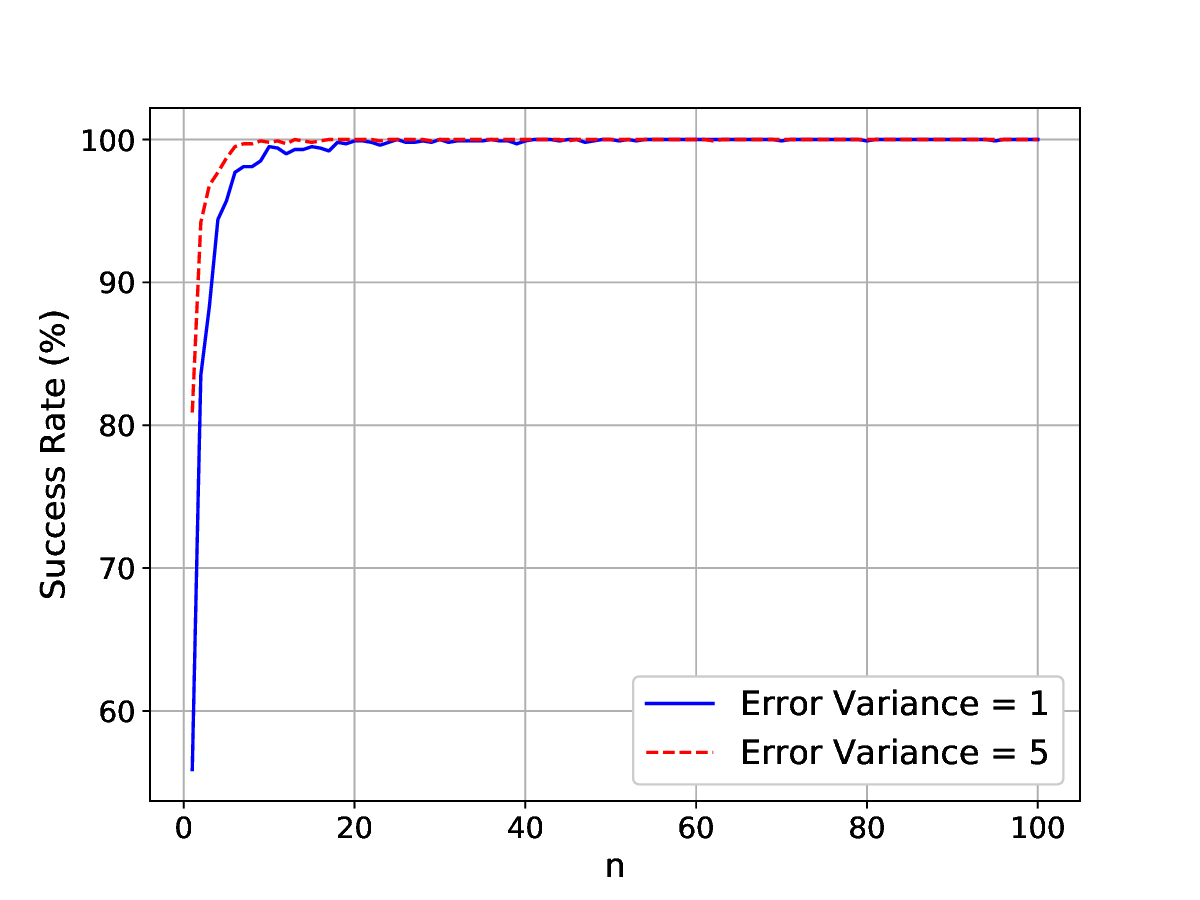}
        \caption{Comparison of the error detection rate with different error variance.}
        \label{fig::error_dection_rate}
    \end{minipage}
    \hfill
    \begin{minipage}[t]{0.48\textwidth}
        \centering
        \includegraphics[scale=0.47]{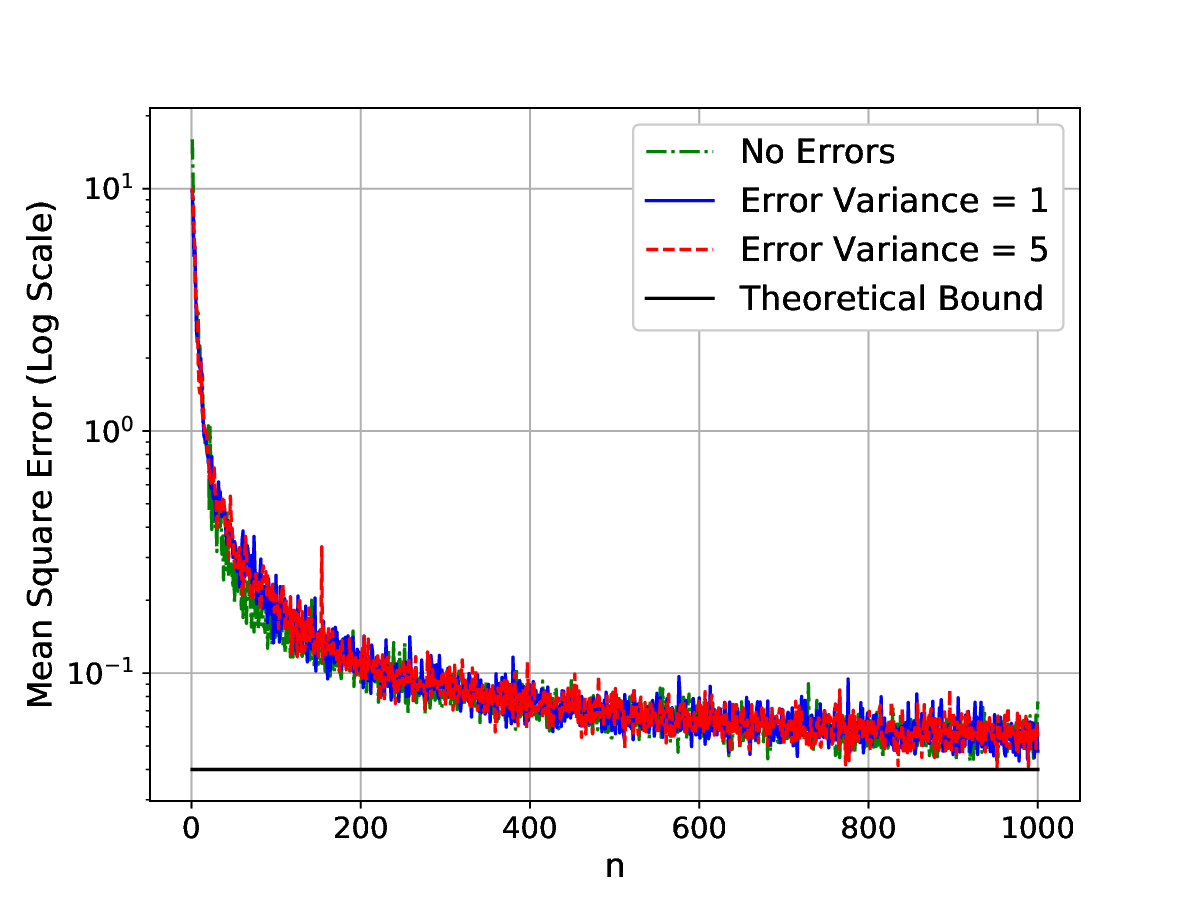}
        \caption{Comparison of the mean square error with different error variance.}
        \label{fig::lmse}
    \end{minipage}
\end{figure}

\section{Numerical Simulations}
{
In this section, we conduct numerical simulations to validate the correctness of the proposed coding scheme.
The simulations consider the setting with $\mathsf{N} = 12$ nodes, including $\mathsf{T} = 5$ colluding nodes, $\mathsf{E} = 2$ erased nodes, and $\mathsf{A} = 2$ adversarial nodes, satisfying $\mathsf{N}=\mathsf{T}+2\mathsf{A}+\mathsf{E}+1$.
We let $\eta = 1$, and $\sigma=0.5$. 
Among the nodes, $\mathsf{E}=2$ nodes are randomly erased, while $\mathsf{A} = 2$ adversarial nodes introduce errors by adding zero-mean Gaussian noise with variance $\mathbb{E}[\sigma^2_e]$. To analyze the worst-case scenario, we assume that erased nodes and adversarial nodes do not overlap.
We select evaluation points $x_i = \cos \left( \frac{(2i-1)\pi}{2\mathsf{N}} \right)$ with $i\in[\mathsf{N}]$.  
The simulations are designed to evaluate system performance as the parameter $n$ increases. To reduce the effects of randomness, 1000 independent trials are conducted for each value of $n$.
The simulations aim to show the system performance with increasing $n$, and 1000 simulations are conducted for each $n$ to reduce the effects of randomness.

Figure \ref{fig::error_dection_rate} illustrates the error detection success rate versus $n$ with error variance $\mathbb{E}[\sigma^2_e] \in \{1,5\}$.
We consider decoding successful if none of the selected messages originate from adversaries.
The success detection rates for both error variances converge to nearly 100\% with increasing $n$, which coincides with the analysis in Section \ref{sec::error_correction}.
Besides, the success rate with $\mathbb{E}[\sigma^2_e]=5$ converges more rapidly compared to that with $\mathbb{E}[\sigma^2_e]=1$.
This phenomenon, different from that in the finite field, arises because errors of larger magnitude are more readily detected, whereas small errors have a less significant impact on the computation result, even if they remain undetected.

Figure~\ref{fig::lmse} illustrates ${\tt MSE}$ under three scenarios: no adversaries (the green dash-dot line), $\mathbb{E}[\sigma^2_e]=1$ (the blue solid line), and $\mathbb{E}[\sigma^2_e]=5$ (the red dotted line), in comparison with the theoretical ${\tt MSE}$ bound established in Theorem~\ref{thm::dp} (black line). 
As $n$ increases, all three curves demonstrate similar trends, with ${\tt MSE}$ converging toward the theoretical bound. These results indicate that the proposed scheme is robust to erasures and adversarial perturbations of varying noise levels, enabling accurate recovery of the multiplication result.
}

\appendices
\section{} \label{apd::distortion}
{
We begin by establishing the properties of the derived error locator polynomial $\tilde{E}(x)$ in Appendix~\ref{sec::E_analysis}, and then use these results to prove Proposition~\ref{prop::main} in Appendix~\ref{sec::algorithm_analysis}.

\subsection{Analysis of Polynomial $\tilde{E}(x)$}\label{sec::E_analysis}
As the derived error locator polynomial $\tilde{E}(x)$ (obtained by solving \eqref{equ::actual_linear_system_general}) is defined as a monic degree-$\mathsf{A}$ polynomial, it can be expressed in the factored form as
\begin{align}
    \tilde{E}(x) = \prod_{j=1}^{\mathsf{A}} (x - a_j),
\end{align}
where the roots $\{a_j\}_{j=1}^\mathsf{A}$ are not necessarily real or distinct, and not necessarily located among the evaluation points. 

Before analyzing the polynomial $\tilde{E}(x)$ over a set of given real evaluation points, it is useful to distinguish between evaluation points that are close to the roots of the polynomial and those that are sufficiently far away. 
We first define the minimal spacing and the maximal spacing between evaluation points, and then use them to define  \emph{safe} and \emph{unsafe} points.

\begin{Definition}[Minimum spacing $D_{\min}$, maximum spacing $D_{\max}$]
Let $\{x_i\}_{i=1}^{\mathsf{N}}$ denote a set of $\mathsf{N}$ distinct real evaluation points. The {minimum spacing} $D_{\min}$ among the evaluation points is defined as
\begin{align}
    D_{\min} \triangleq \min_{1\le i < k \le \mathsf{N}} |x_i - x_k|.
\end{align}
The {maximum spacing} $D_{\max}$ among the evaluation points is defined as
\begin{align}
    D_{\max} \triangleq \max_{1\le i < k \le \mathsf{N}} |x_i - x_k|.
\end{align}
\end{Definition}

\begin{Definition}[Safe and unsafe points]
Let $\tilde{E}(x) = \prod_{j=1}^{\mathsf{A}} (x - a_j)$ be a monic degree-$\mathsf{A}$ polynomial with complex roots $\{a_j\}_{j=1}^{\mathsf{A}}$, and let $\{x_i\}_{i=1}^{\mathsf{N}} $ be a set of $\mathsf{N}$ distinct evaluation points. Let $D_{\min} \triangleq \min_{1\le i < k \le \mathsf{N}} |x_i - x_k|$ denote the minimum spacing between evaluation points. An evaluation point $x_i$ is said to be \emph{unsafe} if there exists a root $a_j$ such that $|x_i - a_j| < D_{\min}/2$. Otherwise, $x_i$ is called \emph{safe}.\footnote{Although the roots $a_j$ may lie in the complex plane, the classification of a real evaluation point $ x_i \in \mathbb{R}$ as either \emph{safe} or \emph{unsafe} remains well-defined, since it depends solely on the Euclidean distance $|x_i - a_j|$ in $\mathbb{C}$.}
\end{Definition}

Given that $\tilde{E}(x)$ is a degree-$\mathsf{A}$ polynomial, it can have at most $\mathsf{A}$ roots (counting multiplicity). Under the assumption that the evaluation points are separated by a minimum distance $D_{\min}$, the number of unsafe points is necessarily limited, since each root can only influence a small neighborhood without overlapping with others. The following proposition formalizes this observation by providing a lower bound on the number of safe evaluation points.

\begin{Proposition}
\label{prop::safe_points}
Let $\{x_i\}_{i=1}^{\mathsf{N}}$ denote a set of $\mathsf{N}$ distinct real evaluation points with minimum spacing $D_{\min}$, and let $\tilde{E}(x) = \prod_{j=1}^{\mathsf{A}} (x - a_j)$ be a degree-$\mathsf{A}$ polynomial with complex roots $\{a_j\}_{j=1}^{\mathsf{A}}$. 
Then, the number of {safe} points is at least $\mathsf{N} - \mathsf{A}$.
\end{Proposition}

\begin{proof}
See Appendix \ref{proof::safe_points}.
\end{proof}

We then establish a lower bound on the magnitude of the error locator polynomial $\tilde{E}(x)$ at a subset of the evaluation points.

\begin{Proposition}\label{prop::nonzero_proposition}
    Let $\{x_i\}_{i=1}^{\mathsf{N}}$ denote a set of $\mathsf{N}$ distinct real evaluation points with minimum spacing $D_{\min}$, and let $\tilde{E}(x) = \prod_{j=1}^{\mathsf{A}} (x - a_j)$ be a degree-$\mathsf{A}$ polynomial with complex roots $\{a_j\}_{j=1}^{\mathsf{A}}$. 
    Then, there exist at least $\mathsf{N} - \mathsf{A}$ safe points and each satisfies $|\tilde{E}(x_{i})| \ge \left( \frac{D_{\min}}{2} \right)^\mathsf{A}$.
\end{Proposition}

\begin{proof}
See Appendix \ref{proof::nonzero_proposition}.
\end{proof}

\begin{Proposition}\label{prop::safe_ratio_bound}
Let $\{x_i\}_{i=1}^{\mathsf{N}}$ denote a set of $\mathsf{N}$ distinct real evaluation points with minimum spacing $D_{\min}$ and maximum spacing $D_{\max}$, and let $\tilde{E}(x) = \prod_{j=1}^{\mathsf{A}} (x - a_j)$ be a degree-$\mathsf{A}$ polynomial with complex roots $\{a_j\}_{j=1}^{\mathsf{A}}$. 
Suppose $x_k$ is a {safe} point, then for any other evaluation point $x_i$,
\begin{align}
    \frac{\left| \tilde{E}(x_i) \right|}{\left| \tilde{E}(x_k) \right|} \le \left( 1+ \frac{2D_{\max}}{D_{\min}} \right)^\mathsf{A}.
\end{align}
\end{Proposition}

\begin{proof}
See Appendix \ref{proof::safe_ratio_bound}.
\end{proof}

\subsection{Analysis of Decoding Algorithm}\label{sec::algorithm_analysis}
Recall that the decoder receives messages $\{Y_{s_i}\}_{i \in [\mathsf{T}+2\mathsf{A}+1]}$ with $\mathcal{S}=\{s_{1}, s_2, \cdots, s_{\mathsf{T}+2\mathsf{A}+1}\}$, and $\mathcal{A} \subseteq \mathcal{S}$ with $|\mathcal{A}| \le \mathsf{A}$ denotes the subset of indices corresponding to adversarial nodes.
Note that the evaluation points $\{x_i\}_{i=1}^{\mathsf{N}}$ are fixed by the coding scheme. 
The set of selected evaluation points is $\{x_{s_i}\}_{i=1}^{\mathsf{T}+2\mathsf{A}+1}$ with the minimum spacing $D_{\min}$ and the maximum spacing $D_{\max}$.
In Algorithm \ref{alg::error_locating}, we derive the coefficients of $\tilde{E}(x)$ and $\tilde{Q}(x)$ by letting  
\begin{align}\label{equ::real_condition}
    Y_{s_i} \tilde{E}(x_{s_i}) =  \tilde{Q}(x_{s_i}),    
\end{align}
with $i\in [\mathsf{T}+2\mathsf{A}+1]$.
Here we partition the above condition according to the two disjoint index sets: adversarial node set $\mathcal{A}$, and non-adversarial node set $\mathcal{S} \setminus \mathcal{A}$. 

For the adversarial node $s_i\in \mathcal{A}$, the decoder imposes the relation
\begin{align}\label{equ::relation_adversarial}
    Y_{s_i} \tilde{E}(x_{s_i}) =  \tilde{Q}(x_{s_i}),
\end{align}
which contributes at most $|\mathcal{A}| \le \mathsf{A}$ equations.

For the non-adversarial node $s_i\in \mathcal{S}\setminus \mathcal{A}$,  \eqref{equ::polynomial_C} leads to 
\begin{align}
    Y_{s_i}=\tilde{C}_{s_i} = P(x_{s_i}) + \frac{r_{s_i}}{n^2} + \frac{t_{s_i}}{n^{\frac{5}{2}}}+\frac{u_{s_i}}{n^{3}},
\end{align}
where $\{r_{s_i}\}_{s_i\in \mathcal{S}\setminus \mathcal{A}}$, $\{t_{s_i}\}_{s_i\in \mathcal{S}\setminus \mathcal{A}}$, and $\{u_{s_i}\}_{s_i\in \mathcal{S}\setminus \mathcal{A}}$ are determined by specific multiplication and are independent of $n$.
The relation in \eqref{equ::real_condition} for non-adversaries can be written as,
\begin{align} \label{equ::relation_nonadversarial}
    \left(P(x_{s_i}) + \frac{r_{s_i}}{n^2} + \frac{t_{s_i}}{n^{\frac{5}{2}}}+\frac{u_{s_i}}{n^{3}}\right) \tilde{E}(x_{s_i}) =  \tilde{Q}(x_{s_i}),
\end{align}
with $s_i\in \mathcal{S}\setminus \mathcal{A}$. The above contributes at least $\mathsf{T}+\mathsf{A}+1$ equations as $|\mathcal{S}\setminus \mathcal{A}| =\mathsf{T}+2\mathsf{A}+1 -|\mathcal{A}| \ge \mathsf{T}+\mathsf{A}+1$.

As $\{r_{s_i}\}_{s_i\in \mathcal{S}\setminus \mathcal{A}}$, $\{t_{s_i}\}_{s_i\in \mathcal{S}\setminus \mathcal{A}}$, and $\{u_{s_i}\}_{s_i\in \mathcal{S}\setminus \mathcal{A}}$ are random variables independent of parameter $n$, the following proposition provides an upper bound on the corresponding tail probability.

\begin{Proposition}\label{prop::tail}
    Let $n>0$, \( \beta > 0 \) be given parameters. For each \( s_i \in \mathcal{S} \setminus \mathcal{A} \), the random variables \( \{r_{s_i}\} \), \( \{t_{s_i}\} \), and \( \{u_{s_i}\} \) satisfy the following probabilistic tail bounds:
    \begin{align}
    \mathbb{P}\left(\left| {r_{s_i}} + \frac{t_{s_i}}{n^{\frac{1}{2}}}+\frac{u_{s_i}}{n} \right| \ge n^{\beta}\right) 
    \le 9n^{-2\beta} \Bigg[
        \sum_{t=2}^{\mathsf{T}} (t-1)x_{s_i}^{2t}
        + \sum_{t=\mathsf{T}+1}^{2\mathsf{T}-2} (2\mathsf{T}-1-t) x_{s_i}^{2t}  + n^{-1} \left( 2\sigma^2 \sum_{t=1}^{\mathsf{T}-1} x_{s_i}^{2(\mathsf{T}+t)} \right)
        + n^{-2} \left( \sigma^2 x_{s_i}^{4\mathsf{T}} \right)
    \Bigg].
\end{align}
\end{Proposition}

\begin{proof}
    See Appendix \ref{proof::tail}.
\end{proof}

For further analysis, we introduce a degree-$(\mathsf{T}+\mathsf{A})$ polynomial $H(x) = \tilde{E}(x) P(x) - \tilde{Q}(x)=\sum_{j=0}^{\mathsf{T}+\mathsf{A}} h_j x^j$.
By evaluating $H(x)$ at non-adversarial positions, we have that for ${s_i}\in \mathcal{S} \setminus \mathcal{A}$,
\begin{align} \label{equ::evaluation_H}
    H(x_{s_i})=\tilde{E}(x_{s_i}) P(x_{s_i}) - \tilde{Q}(x_{s_i})  \overset{(a)}{=} \tilde{E}(x_{s_i}) P(x_{s_i}) - \tilde{E}(x_{s_i}) \left(P(x_{s_i}) + \frac{r_{s_i}}{n^2} + \frac{t_{s_i}}{n^{\frac{5}{2}}}+\frac{u_{s_i}}{n^{3}}\right) = -\tilde{E}(x_{s_i}) \left(\frac{r_{s_i}}{n^2} + \frac{t_{s_i}}{n^{\frac{5}{2}}}+\frac{u_{s_i}}{n^{3}}\right),
\end{align}
where $(a)$ is due to \eqref{equ::relation_nonadversarial}.
Since there are at least $\mathsf{T} + \mathsf{A} + 1$ distinct evaluation points satisfying \eqref{equ::evaluation_H}, and $H(x)$ is a polynomial of degree $\mathsf{T}+\mathsf{A}$, $H(x)$ can be uniquely determined via interpolation based on \(\mathsf{T} + \mathsf{A} + 1\) evaluations.
Let \(\{s'_0, s'_1, \ldots, s'_{\mathsf{T}+\mathsf{A}}\} \subseteq \mathcal{S} \setminus \mathcal{A}\) denote the indices of any such \(\mathsf{T} + \mathsf{A} + 1\) non-adversarial nodes.  
Then, \(H(x)\) can be expressed via Lagrange interpolation as, 
\begin{align}\label{equ::polynomial_H}
    H(x)= \sum_{j=0}^{\mathsf{T}+\mathsf{A}} H(x_{s'_j}) \ell_j(x) \overset{(a)}{=} -\sum_{j=0}^{\mathsf{T}+\mathsf{A}} \left( \tilde{E}(x_{s_j'}) \left(\frac{r_{s_j'}}{n^2} + \frac{t_{s_j'}}{n^{\frac{5}{2}}}+\frac{u_{s_j'}}{n^{3}}\right) \prod_{{0 \le k \le \mathsf{T}+\mathsf{A} ,k \ne j}} \frac{x - x_{s'_k}}{x_{s'_j} - x_{s'_k}} \right),
\end{align}
where $\ell_j(x) = \prod_{{0 \le k \le \mathsf{T}+\mathsf{A} ,k \ne j}} \frac{x - x_{s'_k}}{x_{s'_j} - x_{s'_k}}$, and $(a)$ follows from \eqref{equ::evaluation_H}.
By evaluating $H(x)$ at adversarial points, we have that for $s_i\in \mathcal{A}$,
\begin{align}\label{equ::polynomial_H_adversary}
    H(x_{s_i}) = \tilde{E}(x_{s_i}) P(x_{s_i}) - \tilde{Q}(x_{s_i})  \overset{(a)}{=} \tilde{E}(x_{s_i})\left(P(x_{s_i}) - Y_{s_i} \right), 
\end{align}
where $(a)$ is due to \eqref{equ::relation_adversarial}.
For given distortion $\Delta_{s_i} = |P(x_{s_i}) - Y_{s_i}|$ and based on \eqref{equ::polynomial_H} and \eqref{equ::polynomial_H_adversary}, it follows that, for $s_i \in \mathcal{A}$,
\begin{align} \label{equ::adversary_evaluation_H}
    \left|\tilde{E}(x_{s_i})\right| &= \frac{1}{\Delta_{s_i} n^2}\left| \sum_{j=0}^{\mathsf{T}+\mathsf{A}} \left( \tilde{E}(x_{s_j'}) \left({r_{s_j'}} + \frac{t_{s_j'}}{n^{\frac{1}{2}}}+\frac{u_{s_j'}}{n^{}}\right) \prod_{{0 \le k \le \mathsf{T}+\mathsf{A} ,k \ne j}} \frac{{x_{s_i}} - x_{s'_k}}{x_{s'_j} - x_{s'_k}} \right)\right| \nonumber \\
    &\overset{(a)}{\le} \frac{1}{\Delta_{s_i} n^2} \sum_{j=0}^{\mathsf{T}+\mathsf{A}} \left|  \left({r_{s_j'}} + \frac{t_{s_j'}}{n^{\frac{1}{2}}}+\frac{u_{s_j'}}{n^{}}\right) \prod_{{0 \le k \le \mathsf{T}+\mathsf{A} ,k \ne j}} \frac{{x_{s_i}} - x_{s'_k}}{x_{s'_j} - x_{s'_k}} \right| \left| \tilde{E}(x_{s_j'}) \right| \nonumber \\
    &\overset{(b)}{\le}
    \frac{1}{\Delta_{s_i} n^2} \sum_{j=0}^{\mathsf{T}+\mathsf{A}} \left|  \left({r_{s_j'}} + \frac{t_{s_j'}}{n^{\frac{1}{2}}}+\frac{u_{s_j'}}{n^{}}\right)
    \left( \frac{D_{\max}}{D_{\min}} \right)^{\mathsf{T}+\mathsf{A}} \right| \left| \tilde{E}(x_{s_j'}) \right|
\end{align}
where $(a)$ follows from the triangle inequality,
$(b)$ follows from the definition of $D_{\min}$ and $D_{\max}$.

Based on Proposition \ref{prop::safe_points}, among the $\mathsf{T}+\mathsf{A}+1$ evaluation points $\{x_{s'_j}\}_{j=0}^{\mathsf{T}+\mathsf{A}}$, there exist at least $\mathsf{T}+1$ safe points. Without loss of generality, we assume that evaluation points with indices $\{s'_j\}_{j=0}^\mathsf{T}$ are safe.
Based on Proposition \ref{prop::safe_ratio_bound} and \eqref{equ::adversary_evaluation_H}, we have that for any adversary node $s_i \in \mathcal{A}$ and safe evaluation point $s'_l$ with $l\in \{0,\cdots, \mathsf{T}\}$,
\begin{align}
    \left|\tilde{E}(x_{s_i})\right| \le  \frac{1}{\Delta_{s_i} n^2}
     \left( \frac{D_{\max}}{D_{\min}} \right)^{\mathsf{T}+\mathsf{A}} 
    \left(\left|  {r_{s_l'}} + \frac{t_{s_l'}}{n^{\frac{1}{2}}}+\frac{u_{s_l'}}{n^{}} \right| + \left( 1+ \frac{2D_{\max}}{D_{\min}} \right)^\mathsf{A} \sum_{j=0, j\neq l}^{\mathsf{T}+\mathsf{A}} \left|  {r_{s_j'}} + \frac{t_{s_j'}}{n^{\frac{1}{2}}}+\frac{u_{s_j'}}{n^{}}  \right| \right)
     \left| \tilde{E}(x_{s_l'}) \right|.
\end{align}

As Proposition \ref{prop::nonzero_proposition} implies that $\left|\tilde{E}(x_{s_l'})\right| \neq 0$, if 
$$
\frac{\left|\tilde{E}(x_{s_i})\right|}{\left| \tilde{E}(x_{s_l'}) \right|}=
\frac{1}{\Delta_{s_i} n^2}
     \left( \frac{D_{\max}}{D_{\min}} \right)^{\mathsf{T}+\mathsf{A}} 
    \left(\left|  {r_{s_l'}} + \frac{t_{s_l'}}{n^{\frac{1}{2}}}+\frac{u_{s_l'}}{n^{}} \right| + \left( 1+ \frac{2D_{\max}}{D_{\min}} \right)^\mathsf{A} \sum_{j=0, j\neq l}^{\mathsf{T}+\mathsf{A}} \left|  {r_{s_j'}} + \frac{t_{s_j'}}{n^{\frac{1}{2}}}+\frac{u_{s_j'}}{n^{}}  \right| \right) < 1,$$ it follows that 
$\left|\tilde{E}(x_{s_i})\right|< \left| \tilde{E}(x_{s_l'}) \right|$ for every safe point $x_{s'_l}$.
For the adversary node $s_i\in \mathcal{A}$ satisfies the distortion $\Delta_{s_i} \ge n^{-1.6}$, the above inequality can be written as
\begin{align}
\label{equ::detection_condition}
    n^{-0.4} \left( \frac{D_{\max}}{D_{\min}} \right)^{\mathsf{T}+\mathsf{A}} 
    \left(\left|  {r_{s_l'}} + \frac{t_{s_l'}}{n^{\frac{1}{2}}}+\frac{u_{s_l'}}{n^{}} \right| + \left( 1+ \frac{2D_{\max}}{D_{\min}} \right)^\mathsf{A} \sum_{j=0, j\neq l}^{\mathsf{T}+\mathsf{A}} \left|  {r_{s_j'}} + \frac{t_{s_j'}}{n^{\frac{1}{2}}}+\frac{u_{s_j'}}{n^{}}  \right| \right) < 1
\end{align}

According to Proposition \ref{prop::tail}, with high probability (specifically at least $1 - \delta(n)$ where $\delta(n) = O(n^{-2\beta})$) the term $\left| r_{s_i} + \frac{t_{s_i}}{n^{1/2}} + \frac{u_{s_i}}{n} \right|$ is upper bounded  by $n^\beta$. 
By setting $\beta = 0.05$, it follows that 
\begin{align}
\label{equ::high_prob}
    \Pr\left[\left| {r_{s_i}} + \frac{t_{s_i}}{n^{\frac{1}{2}}}+\frac{u_{s_i}}{n} \right| < n^{0.05}\right] 
    > 1- 9n^{-0.1} \Bigg[
        \sum_{t=2}^{\mathsf{T}} (t-1)x_{s_i}^{2t}
        + \sum_{t=\mathsf{T}+1}^{2\mathsf{T}-2} (2\mathsf{T}-1-t) x_{s_i}^{2t}  + n^{-1} \left( 2\sigma^2 \sum_{t=1}^{\mathsf{T}-1} x_{s_i}^{2(\mathsf{T}+t)} \right)
        + n^{-2} \left( \sigma^2 x_{s_i}^{4\mathsf{T}} \right)
    \Bigg].
\end{align}
Substituting the bound $n^{0.05}$ into \eqref{equ::detection_condition}, we obtain
\begin{align}\label{equ::n0}
    1& > n^{-0.4} \left( \frac{D_{\max}}{D_{\min}} \right)^{\mathsf{T}+\mathsf{A}} 
    \left( n^{0.05} + \left( 1+ \frac{2D_{\max}}{D_{\min}} \right)^\mathsf{A} {\left(\mathsf{T}+\mathsf{A}\right)} n^{0.05} \right) \nonumber \\
    n & > \left( \left( \frac{D_{\max}}{D_{\min}} \right)^{\mathsf{T}+\mathsf{A}} 
    \left[ 1 + \left( 1+ \frac{2D_{\max}}{D_{\min}} \right)^\mathsf{A} (\mathsf{T}+\mathsf{A}) \right] \right)^{\frac{20}{7}} = n_0
\end{align}

Consequently, when $n > n_0$, we have $\left|\tilde{E}(x_{s_i})\right| < \left|\tilde{E}(x_{s_l'})\right|$ with probability approaching 1, where $x_{s_l'}$ corresponds to a safe point.
As Proposition \ref{prop::safe_points} implies that there exist at least $\mathsf{T}+1$ safe points among the $\mathsf{T}+\mathsf{A}+1$ evaluation points $\{x_{s'_j}\}_{j=0}^{\mathsf{T}+\mathsf{A}}$, for adversary node $s_i\in \mathcal{A}$ with distortion $\Delta_{s_i} \ge n^{-1.6}$, the corresponding $\left|\tilde{E}(x_{s_i})\right|$ is smaller than the value evaluated at least $\mathsf{T}+1$ safe points.
Therefore Algorithm \ref{alg::error_locating} can exclude all adversaries $s_i\in \mathcal{A}$  satisfying the distortion $\Delta_{s_i} \ge n^{-1.6}$ with high probability.

Hence, Proposition \ref{prop::main} is proved.

\section{Additional Proofs}

\subsection{Proof of Proposition \ref{prop::safe_points}}
\label{proof::safe_points}
Consider $\mathsf{N}$ distinct real evaluation points $\{x_i\}_{i=1}^{\mathsf{N}}$ with the minimal spacing $D_{\min}$.
We now establish a key fact: each root $a_j$ can cause at most one evaluation point to be unsafe. Indeed, suppose two evaluation points $x_i$ and $x_k$ both satisfy $|x_i - a_j| < D_{\min}/2$ and $|x_k - a_j| < D_{\min}/2$ for the same $a_j$. Then by the triangle inequality,
\begin{align}
    |x_i - x_k| \le |x_i - a_j| + |x_k - a_j| < \frac{D_{\min}}{2} + \frac{D_{\min}}{2} = D_{\min},
\end{align}
contradicting the minimal spacing condition among $\{x_i\}_{i=1}^\mathsf{N}$. Hence, each root corresponds to at most one unsafe evaluation point, implying that the total number of unsafe points is at most $\mathsf{A}$. Consequently, at least $\mathsf{N}-\mathsf{A}$ of the $\mathsf{N}$ evaluation points must be safe.
Hence Proposition \ref{prop::safe_points} is proved.

\subsection{Proof of Proposition \ref{prop::nonzero_proposition}} \label{proof::nonzero_proposition}
Based on Proposition \ref{prop::safe_points}, we have that at least $\mathsf{N}-\mathsf{A}$ of the $\mathsf{N}$ evaluation points must be safe.
Next, consider any such safe point $x_i$. By definition of safeness, we have $|x_i - a_j| \ge D_{\min}/2$ for every $j \in [\mathsf{A}]$. Thus $|\tilde{E}(x_i)|$ is lower bounded by
\begin{align}
    |\tilde{E}(x_i)| = \prod_{j=1}^\mathsf{A} |x_i - a_j| \ge \left( \frac{D_{\min}}{2} \right)^\mathsf{A}.
\end{align}
Hence, Proposition \ref{prop::nonzero_proposition} is proved.

\subsection{Proof of Proposition \ref{prop::safe_ratio_bound}}
 \label{proof::safe_ratio_bound}
 For safe points $x_k$ and any other points $x_i$, as Proposition \ref{prop::nonzero_proposition} implies that $|\tilde{E}(x_k)| \neq 0$, the ratio between $|\tilde{E}(x_i)|$ and $|\tilde{E}(x_k)|$ can be written as 
 \begin{align}
     \frac{\left| \tilde{E}(x_i) \right|}{\left| \tilde{E}(x_k) \right|} &\overset{(a)}{\le} \prod_{j=1}^{\mathsf{A}} \frac{\left| x_i - a_j \right|}{\left| x_k - a_j \right|} \nonumber \\
     &\overset{(b)}{\le} \prod_{j=1}^{\mathsf{A}} \frac{\left| x_i - x_k \right| + \left| x_k - a_j \right|}{\left| x_k - a_j \right|} \nonumber \\
     &= \prod_{j=1}^{\mathsf{A}} \left(1+ \frac{\left| x_i - x_k \right|}{\left| x_k - a_j \right|} \right)
      \nonumber \\
     &\overset{(c)}{\le} \prod_{j=1}^{\mathsf{A}} \left(1+ \frac{D_{\max}}{D_{\min}/2} \right) \nonumber \\
     &=\left( 1+ \frac{2D_{\max}}{D_{\min}} \right)^\mathsf{A},
 \end{align}
where $(a)$ follows from the factorized expression of $\tilde{E}(x)$, $(b)$ follows form the triangle inequality, $(c)$ follows from the fact that $x_k$ is a safe points and the definition of maximal spacing $D_{\max}$.

\subsection{Proof of Proposition \ref{prop::tail}} \label{proof::tail}

From \eqref{equ::encoding_poly}, we have that each of $\{r_{s_i}\}_{s_i\in \mathcal{S}\setminus \mathcal{A}}$, $\{t_{s_i}\}_{s_i\in \mathcal{S}\setminus \mathcal{A}}$, and $\{u_{s_i}\}_{s_i\in \mathcal{S}\setminus \mathcal{A}}$ can be explicitly written as
\begin{subequations}
    \begin{align}
        &r_{s_i}= \left( \sum_{t=1}^{\mathsf{T}-1} R_{t+1} x_{s_i}^t \right) \cdot \left( \sum_{t=1}^{\mathsf{T}-1} S_{t+1} x_{s_i}^t \right)= \sum_{t=2}^{2\mathsf{T}-2} \left( \sum_{j=1, 1 \le t - j \le \mathsf{T} - 1}^{\mathsf{T}-1} R_{j+1} S_{t - j + 1} \right) x_{s_i}^t,\\
        &t_{s_i}=\sum_{t=1}^{\mathsf{T}-1} \left( R_{t+1} S_1 + R_1 S_{t+1} \right) x_{s_i}^{\mathsf{T} + t},\\
        &u_{s_i}= R_1 S_1 x_{s_i}^{2\mathsf{T}}.
    \end{align}
\end{subequations}

As $\{R_t, S_t\}_{t=1}^{\mathsf{T}}$ are independent and zero-mean random variables, and $\mathbb{E}[R_1^2]=\mathbb{E}[S_1^2]=\sigma^2$, $\mathbb{E}[R_t^2]=\mathbb{E}[S_t^2]=1$ for $t\in\{2,\cdots, \mathsf{T}\}$, it follows that
\begin{subequations}
    \begin{align}
        &\mathbb{E}[r_{s_i}^2] = \sum_{t=2}^{\mathsf{T}} (t-1)x_{s_i}^{2t}+ \sum_{t=\mathsf{T}+1}^{2\mathsf{T}-2} (2\mathsf{T}-1-t) x_{s_i}^{2t},\\
        &\mathbb{E}[t_{s_i}^2] = 2 \sigma^2 \sum_{t=1}^{\mathsf{T}-1} x_{s_i}^{2(\mathsf{T} + t)},\\
        &\mathbb{E}[u_{s_i}^2] = \sigma^2 x_{s_i}^{4\mathsf{T}},
    \end{align}
\end{subequations}
and $\mathbb{E}[r_{s_i}]=\mathbb{E}[t_{s_i}] = \mathbb{E}[u_{s_i}]=0.$
Hence, by utilizing  Chebyshev's inequality, we have that for $\beta>0$ and $n$,
\begin{subequations}
    \begin{align}
        &\mathbb{P}\left( |r_{s_i}| \ge n^\beta \right) \le {n}^{-2\beta} \left( \sum_{t=2}^{\mathsf{T}} (t-1)x_{s_i}^{2t}+ \sum_{t=\mathsf{T}+1}^{2\mathsf{T}-2} (2\mathsf{T}-1-t) x_{s_i}^{2t} \right), \\
        &\mathbb{P}\left( |t_{s_i}| \ge n^\beta \right) \le {n}^{-2\beta} \left( 2 \sigma^2 \sum_{t=1}^{\mathsf{T}-1} x_{s_i}^{2(\mathsf{T} + t)} \right), \\
        &\mathbb{P}\left( |u_{s_i}| \ge n^\beta \right) \le n^{-2\beta} \left(  \sigma^2 x_{s_i}^{4\mathsf{T}} \right).
    \end{align}
\end{subequations}

Based on the following and the union bound, it follows that
\begin{align}
    \mathbb{P}\left(\left| {r_{s_i}} + \frac{t_{s_i}}{n^{\frac{1}{2}}}+\frac{u_{s_i}}{n} \right| \ge n^{\beta}\right) &\le \mathbb{P}\left( \left|r_{s_i}\right| \ge \frac{1}{3} n^\beta \right) + \mathbb{P}\left( \left|t_{s_i}\right| \ge \frac{1}{3} n^{\beta+\frac{1}{2}} \right) + \mathbb{P}\left( \left|u_{s_i}\right| \ge \frac{1}{3} n^{\beta+1} \right) \nonumber \\
    & \le 9n^{-2\beta} \Bigg[
        \sum_{t=2}^{\mathsf{T}} (t-1)x_{s_i}^{2t}
        + \sum_{t=\mathsf{T}+1}^{2\mathsf{T}-2} (2\mathsf{T}-1-t) x_{s_i}^{2t}  + n^{-1} \left( 2\sigma^2 \sum_{t=1}^{\mathsf{T}-1} x_{s_i}^{2(\mathsf{T}+t)} \right)
        + n^{-2} \left( \sigma^2 x_{s_i}^{4\mathsf{T}} \right)
    \Bigg].
\end{align}

Hence, Proposition \ref{prop::tail} is then proved.

}

\bibliographystyle{IEEEtran.bst}
\bibliography{main}

\end{document}